\documentclass[a4paper, amsfonts, amssymb, amsmath, reprint, showkeys, nofootinbib, twoside, floatfix, pre,superscriptaddress]{revtex4-2}

\usepackage[utf8]{inputenc}
\usepackage[left=23mm,right=13mm,top=35mm,columnsep=15pt]{geometry} 
\usepackage{amsmath}
\usepackage{amssymb}
\usepackage{esint}
\usepackage[unicode=true, pdfusetitle, bookmarks=true, bookmarksnumbered=false, bookmarksopen=false, breaklinks=false, pdfborder={0 0 1}, backref=false, colorlinks=false]{hyperref}
\usepackage{graphicx}
\graphicspath{ {./images/} }
\usepackage{xcolor}
\usepackage[shortlabels]{enumitem}
\makeatletter
\usepackage{makeidx}\usepackage{amsfonts}

\makeatother

\begin{document}
\bibliographystyle{myunsrt}

\title{Interacting Copies of Random Constraint Satisfaction Problems}

\author{Maria Chiara Angelini}
\affiliation{Dipartimento di Fisica, Sapienza Università di Roma, Piazzale Aldo Moro 5, Rome 00185, Italy} 
\affiliation{CNR-Nanotec, Rome unit and INFN, sezione di Roma1, Piazzale Aldo Moro 5, Rome 00185, Italy}
\author{Louise Budzynski}
\affiliation{DI ENS, École Normale Supérieure, PSL, CNRS, INRIA}
\affiliation{Dipartimento di Fisica, Sapienza Università di Roma, Piazzale Aldo Moro 5, Rome 00185, Italy} 
\author{Federico Ricci-Tersenghi}
\affiliation{Dipartimento di Fisica, Sapienza Università di Roma, Piazzale Aldo Moro 5, Rome 00185, Italy} 
\affiliation{CNR-Nanotec, Rome unit and INFN, sezione di Roma1, Piazzale Aldo Moro 5, Rome 00185, Italy}

\newcommand{\com}{\textcolor{blue}}

\begin{abstract}
We study a system of $y=2$ coupled copies of a well-known constraint satisfaction problem (random hypergraph bicoloring) to examine how the ferromagnetic coupling between the copies affects the properties of the solution space. We solve the replicated model by applying the cavity method to the supervariables taking $2^y$ values.
Our results show that a coupling of strength $\gamma$ between the copies decreases the clustering threshold $\alpha_d(\gamma)$, at which typical solutions shatters into disconnected components, therefore preventing numerical methods such as Monte Carlo Markov Chains from reaching equilibrium in polynomial time. 
This result needs to be reconciled with the observation that, in models with coupled copies, denser regions of the solution space should be more accessible.
Additionally, we observe a change in the nature of the clustering phase transition, from discontinuous to continuous, in a wide $\gamma$ range. 
We investigate how the coupling affects the behavior of the Belief Propagation (BP) algorithm on finite-size instances and find that BP convergence is significantly impacted by the continuous transition. 
These results highlight the importance of better understanding algorithmic performance at the clustering transition, and call for a further exploration into the optimal use of re-weighting strategies designed to enhance algorithmic performances.
\end{abstract}

\maketitle

\section{Introduction}
Combinatorial optimization problems are widespread in real life as well as in many scientific disciplines: from physics, in the computation of ground-state configurations, to statistical inference with likelihood maximization, and in many areas of computer science.
Among the many types of combinatorial optimization problems, random Constraint Satisfaction Problems (CSPs) stand out as ideal prototypical problems for studying the average-case hardness of algorithms. Prominent examples of CSPs include the $q$-coloring problem on graphs, and the Boolean satisfiability problem. 
In an instance of CSP, a set of $N$ variables is subjected to $M$ constraints, and the decision version of this problem consists in finding an assignment to the variables satisfying all constraints simultaneously.

The average-case hardness of a CSP can be analyzed by introducing random ensembles of instances.
Thanks to a formal analogy between CSPs and spin-glasses, the application of methods coming from statistical physics of disordered systems, such as the replica and the cavity method, has led to a detailed description of the solution space of random instances \cite{MonassonZecchina99b, BiroliMonasson00, MezardParisi02, MertensMezard06, krzakala2007gibbs, mezard2009information}. 
Many of these predictions were later proven rigorously \cite{AchlioptasRicci06, AchlioptasCoja-Oghlan08, molloy_col_freezing, ding2014proof}.
In this context, a particularly interesting regime is the large size (or thermodynamic) limit, where both the number of constraints $M$ and the number of variables $N$ are sent to $\infty$, at a fixed ratio $\alpha=M/N$.
Random CSPs exhibit threshold phenomena (or phase transitions) in this limit, as $\alpha$ increases. 
The most prominent of these phase transitions is the SAT/UNSAT one, at $\alpha_{\rm sat}$, above which no solutions exists with high probability.
In the satisfiable phase $\alpha<\alpha_{\rm sat}$, many other phase transitions occur, affecting the geometrical structure of the solution set.

One could hope that this detailed description can shed light on the average-case algorithmic hardness of CSPs, helping understanding the behavior of algorithms in the satisfiable phase. 
However, many of these phase transitions affect the equilibrium properties of the solution-set, making difficult the connection with algorithms -- working mostly in the out-of-equilibrium regime (either because they do not satisfy detailed balance, or because they are run on time scales shorter than their relaxation time).

Recently, the line of work on the Overlap Gap Property \cite{Gamarnik_21, GaSu17, CoHaHe17, BrHu22, gamarnik2025turing} could rule out a large class of {\it stable} algorithms, for problems exhibiting a strong form of topological discontinuity in the set of distances between near optimal solutions.
This approach has the advantage of describing the properties of all solutions, and not only the typical ones (dominating the uniform measure over the solution-set).
Hence, it does not suffer from the aforementioned discrepancy between the geometrical structure of {\it equilibrium} and {\it out-of-equilibrium} solutions.
On the other hand, widely used algorithms such as Simulated Annealing (SA) \cite{KirkpatrickGelatt83, AnRi23}, and Belief-Propagation guided decimation \cite{MoRiSe07, Ricci-Tersenghi_2009} might fall out of the class of {\it stable} algorithms when used in the regime such that the number of iterations scale quadratically in the system size \cite{angelini2025algorithmic}.

In this paper, we pay particular attention to the clustering (or dynamic) threshold $\alpha_d$, above which the set of solutions splits into a large number of distinct groups of solutions, called clusters, which are internally well connected but well separated from each other.
This transition is also manifested by the appearance of a specific form of long-range correlations between variables, called point-to-set correlations, in the probability distribution defined as the uniform measure over the set of solutions. These correlations prohibit the rapid equilibration of stochastic processes that satisfy the detailed balance condition \cite{MoSe06}, which justifies the alternative name “dynamic” for the clustering transition.
The clustering threshold $\alpha_d$ hence gives a lower bound to the algorithmic threshold $\alpha_{\rm alg}$ above which no algorithm can find solutions to a typical instance, given that below $\alpha_d$ Monte Carlo based algorithms can sample uniformly the solution-set in polynomial time down to arbitrarily small temperatures.
Of course, the lower bound is generally not tight, as e.g., SA is able to find solutions non-uniformly even before reaching equilibrium \cite{BuRiSe19}.

Many structural phase transitions occurring in the satisfiable regime, and in particular the clustering threshold $\alpha_d$, depend on correlations between variables defined over a specific probability distribution over the configuration space, namely the uniform measure across solutions.
In a series of works \cite{BrDaSeZd16, BaInLuSaZe15_long, BaBo16, MaSeSeZa18, BuRiSe19, BuSe20, ZhZh20}, it was demonstrated that introducing a re-weighting of the solution set could significantly move the location of these structural phase transitions, and that this strategy could be used to improve the performance of algorithms searching for a solution to a random CSP instance.

Following this line of thought, in this paper, we study a model of coupled copies (or real replicas) of a CSP instance, where the coupling strength between copies is an external parameter that can be varied to re-weight the solution space.
The model with coupled replicas has a deep connection with the local-entropy approach \cite{BaInLuSaZe15_long} counting the number of solutions in a given neighborhood: once marginalized over the other copies of the model, the resulting re-weighted probability distribution over a single replica favors solutions living in dense regions (i.e., having a large local entropy). 

Our investigation builds upon recent works exploring the effects of coupled copies of CSPs in the planted $q$-coloring on graphs \cite{AnRi23}, and in the binary perceptron \cite{CaDeSe24} in the teacher-student scenario, where it was shown that algorithms sampling from the system of interacting copies (such as replicated SA) outperforms the classical approach using a single copy of the CSP instance when one wants to infer the planted known solution.
Another motivation for studying a system of interacting copies is that it induces long-range interactions between variables of a single copy (once marginalized over the other ones), that could be tuned to decrease the long-range point-to-set correlation between distant variables that appears in the clustered phase. 
It was indeed observed that extending the range of interactions between variables in the re-weighted distribution over the solution-set \cite{BuRiSe19, BuSe20} can delay the clustering threshold $\alpha_d$ to higher values, therefore calling for a generic strategy to extend the interaction range.

In this work, we concentrate our efforts on the bi-coloring problem on random $k$-hypergraphs \cite{CaNaRiZe03}, in which the variables can take two values, and each constraint acts on a $k$-uplet of variables, forbidding monochromatic configurations.
We introduce a model of $y$ copies of the same instance, interacting site-by-site through a ferromagnetic coupling of strength $\gamma$, and examine how the dynamical phase transition behaves as a function of $\gamma$. 
Remarkably, we find that turning on the interaction between copies has the effect of \emph{decreasing} the dynamical threshold $\alpha_d(\gamma)$, effectively shrinking the region where algorithms such as SA and message-passing algorithms can sample instances in polynomial time. 
In addition to the characterization of the phase diagram, we also provide a numerical study of the effect of the coupling strategy on the Belief Propagation algorithm run on finite size graph instances, confirming that in the coupled model the region of convergence of BP get reduced.

These results are surprising, and challenge prior conjectures about the benefits of introducing interacting copies of a system in optimization problems.
They open the door to further investigations into the optimal use of re-weighting strategies.
However, we will also discuss the possible beneficial effects of the coupling that, modifying a discontinuous transition into a continuous one, can  make solution in the clustered phase easier to approximate.

While in the present work the analysis is limited to the case of two copies, $y=2$, our study can be straightforwardly extended to larger values of $y$.

\section{Set-up of the Problem}  

\subsection{Definition of the model}

\subsubsection{The $k$-uniform hyper-graph bi-coloring problem}
\begin{figure}[h!]
	\centering
	\includegraphics[width=0.3\textwidth]{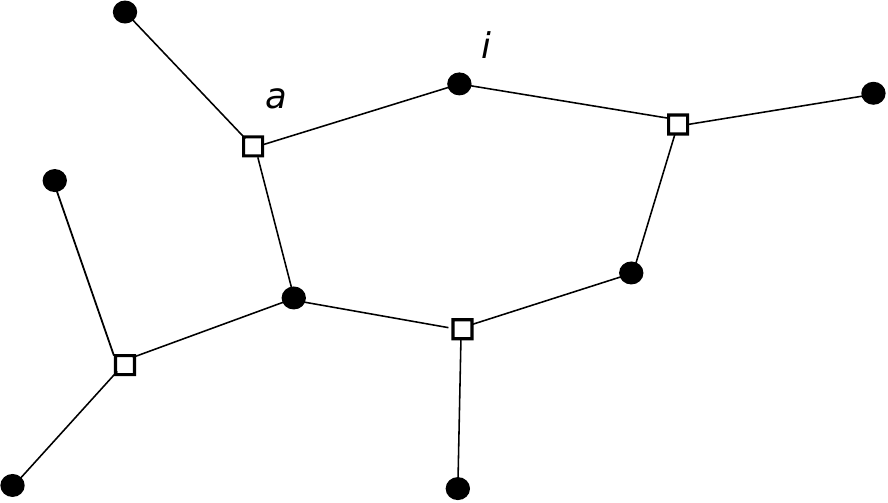}(a)	\\
	\includegraphics[width=0.3\textwidth]{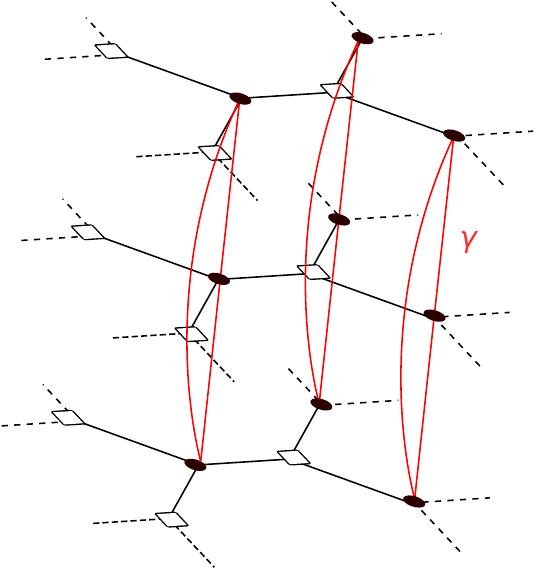}(b)
	\caption{Top (a): An instance of the $k$-uniform hyper-graph bi-coloring problem with $k=3$, $N=8$, $M=4$. Vertices are represented by black circles, hyper-edges by white squares. An edge is drawn between the $a$-th hyper-edge and the vertex $i$ if and only if $i$ is linked to $a$ ($i\in\partial a$).\\
		Bottom (b): A system of $y=3$ interacting copies. Each pair of copies interact site-by-site with a coupling strength $\gamma$ (red edges).}
	\label{fig:bicol_hypergraph}
\end{figure}

We consider in this paper the $k$-uniform hyper-graph bi-coloring problem \cite{CaNaRiZe03}. An instance of this Constraint Satisfaction Problem (CSP) is defined by an hyper-graph $G=(V,E)$, with a set $V$ of $N$ vertices, and a set $E$ of $M$ hyper-edges, each hyper-edge involving a subset of $k$ vertices (see Figure \ref{fig:bicol_hypergraph}, top panel (a)).
A set of $N$ spin variables $\sigma_1,\dots,\sigma_N$ live on the vertices of the graph, with $\sigma_i\in\{-1,1\}$. 
We denote by $\underline{\sigma}=(\sigma_1,\dots,\sigma_N)$ the global configuration of the variables, and $\underline{\sigma}_S$ the configuration of a subset $S\subseteq V$ of the vertices.
A constraint (or clause) is associated to each hyper-edge. The $a$-th constraint is satisfied if and only if there is at least one $+1$ and one $-1$ among the $k$ variables of $\underline{\sigma}_{\partial a}$, where $\partial a$ is the set of vertices contained in an hyper-edge $a\in E$ (and similarly $\partial i$ the set of hyper-edges adjacent to vertex $i\in V$).
A configuration $\underline{\sigma}$ is a solution of the CSP if and only if it satisfies all the $M$ constraints simultaneously.

A convenient way of studying the solution-set $\mathcal{S}(G)$  of a given instance of this problem is to introduce the uniform probability measure:
\begin{align}
	\label{eq:prob_bicol}
	\mu(\underline{\sigma}) = \frac{1}{Z(G)}\prod_{a=1}^M \omega(\underline{\sigma}_{\partial a})
\end{align}
where the normalization factor $Z(G)=|\mathcal{S}(G)|$ counts the number of solutions, and the function $\omega(\sigma_1,\dots,\sigma_k)$ is the indicator function of the event ``the $k$ variables $\sigma_1,\dots,\sigma_k$ are not all equal''.

\subsubsection{Interacting copies}

In this paper, we introduce $y$ copies of a given instance $G=(V,E)$ of the bicoloring problem. 
The system is represented in Figure \ref{fig:bicol_hypergraph}, bottom panel (b). Each pair of copies is interacting site-by site, with a coupling strength $\gamma$.
The probability measure representing this system is:
\begin{align}
	\label{eq:prob_copies}
	\mu_y(\underline{\sigma}^1,\dots,\underline{\sigma}^y) = \frac{1}{Z_y}\prod_{a=1}^M\left(\prod_{s=1}^y\omega(\underline{\sigma}_{\partial a}^s)\right)\prod_{i=1}^Ne^{\frac{\gamma}{2y}\sum_{s\neq t}\sigma_i^s\sigma_i^t}
\end{align}
where the spin variables associated with the copy $s\in\{1,\dots,y\}$ are denoted $\underline{\sigma}^s=(\sigma_1^s,\dots,\sigma_N^s)$. 
A ferromagnetic coupling $\gamma>0$ favors copies $\underline{\sigma}^1,\dots,\underline{\sigma}^y$ in similar configurations, while an anti-ferromagnetic coupling favors distant configurations.

The case of a single copy ($y=1$) is retrieved from (\ref{eq:prob_copies}) in the case of independent copies (setting $\gamma=0$), or when the coupling forces the copies to be identical (sending $\gamma\to\infty$).

Our aim is to study the properties of $\mu_y$ for typical hyper-graphs, by studying random instances. In this paper, we concentrate our efforts on Erd\"os R\'enyi (ER) random hyper-graphs.
An instance of this problem is generated by drawing, independently for each hyper-edges $a\in\{1,\dots,M\}$, the set of adjacent vertices $\partial a$ uniformly at random among the $\binom{N}{k}$ possible $k$-uplets.
We will be interested in the large size (thermodynamic) limit, where both $N$ and $M$ go to infinity at a fixed ratio $\alpha=M/N$ (called the density of constraints).
In this limit, ER random hyper-graphs converge locally to hyper-trees, and the degree distribution follows a Poisson law of parameter $\alpha k$.

\subsection{Belief-Propagation}
In order to study the typical properties of the measure $\mu_y$ (\ref{eq:prob_copies}), we use the cavity method \cite{MezardParisi01,MezardParisi03}, a method efficient on interacting particle models defined on random sparse structures.
The first step of the cavity method amounts to study the model (\ref{eq:prob_copies}) on finite trees, where an exact description of $\mu_y$ in terms of marginals and of the free energy $\ln Z_y$, can be obtained with Belief Propagation (BP).

\subsubsection{Super-spin variables}
The factor graph (Fig.~\ref{fig:bicol_hypergraph} bottom panel (b)) represents the interactions between variables in the measure $\mu_y$.
It contains small loops due to the on-site interaction term $e^{\frac{\gamma}{2y}\sum_{s\neq t}\sigma_i^s\sigma_i^t}$ (represented by the red edges on the figure) on each vertex $i\in V$  of the hyper-graph $G=(V,E)$.
These small loops forbid a direct use of the cavity method -- which is well-suited for tree-like problems --  on this factor graph. 
A natural strategy to circumvent this difficulty is to define super-spin variables $X_i=(\sigma_i^1,\dots,\sigma_i^y)$ on each vertex $i\in V$.
The associated probability measure now writes:
\begin{align}
	\label{eq:prob_superspins}
	\begin{aligned}
		\mu_y(\underline{X})&=\frac{1}{Z_y}\prod_{a=1}^M\Omega(\underline{X}_{\partial a})\prod_{i=1}^N\phi(X_i) \quad  \text{with}\\
		\Omega(\underline{X}_{\partial a})&=\prod_{s=1}^y\omega(\underline{\sigma}_{\partial a}^s) \quad , \quad \phi(X_i)=e^{\frac{\gamma}{2y}\sum_{s\neq t}\sigma_i^s\sigma_i^t} \quad ,
	\end{aligned}
\end{align}
and its associated factor graph is the original hyper-graph $G=(V,E)$.
\subsubsection{Belief Propagation Equations}
For each $i\in V$, $a\in\partial i$, we introduce the variable-to-factor and factor-to-variable BP messages $\eta_{i\to a}$ and $\widehat{\eta}_{a\to i}$, as the marginal probability laws of $X_i$ in the amputated graph where some interactions have been discarded: $\eta_{i\to a}$ is the marginal of $X_i$ when $a$ has been removed, and $\widehat{\eta}_{a\to i}$ is the marginal of $X_i$ when one removes all hyper-edges in $\partial i\setminus a$.
The BP messages obey the following set of equations:
\begin{align}
	\label{eq:BPeqn}
	\begin{aligned}
		\eta_{i\to a}(X_i)&=\frac{\phi(X_i)}{z_{i\to a}}\prod_{b\in\partial i\setminus a}\widehat{\eta}_{b\to i}(X_i) \\
		\widehat{\eta}_{a\to i}(X_i) &= \frac{1}{\widehat{z}_{a\to i}}\sum_{\underline{X}_{\partial a\setminus i}}\Omega(\underline{X}_{\partial a})\prod_{j\in\partial a \setminus i}\eta_{j\to a}(X_j)
	\end{aligned}
\end{align}
where $z_{i\to a}$ and $\widehat{z}_{a\to i}$ are normalization factors.
One can compute the marginal probability of the variable $X_i$ from the set of incoming messages $\{\widehat{\eta}_{a\to i}\}$:
\begin{align}
	\label{eq:BPmarg}
	\mu_i(X_i)  = \frac{\phi(X_i)}{z_i}\prod_{a\in \partial i}\widehat{\eta}_{a\to i}(X_i)
\end{align}
Note that the size of the super-spin variable $X_i=(\sigma_i^1,\dots,\sigma_i^y)$ grows exponentially with the number of copies, which represent a limitation for a numerical representation of the BP messages with a large number of copies. For this reason in the following we will concentrate our attention to the case $y=2$.

\subsection{Ensemble average with the cavity method}
The BP equations (\ref{eq:BPeqn}) are exact when the hyper-graph $G$ is an hyper-tree, and can be used heuristically on any factor graph, even in the presence of loops. 
The message-passing iterative algorithm searching for fixed point of these equations is called Belief Propagation algorithm.
The cavity method is based on the application of Belief Propagation on random hyper-graphs which are locally tree-like in the thermodynamic limit (such as ER random hyper-graphs).

\subsubsection{Replica Symmetric cavity method}
There are different versions of the cavity method, that rely on self-consistent hypothesis on the effect of the long loops that are present in random graphs.
The simplest version, called Replica Symmetric (RS), assumes a fast decay of the correlations between distant variables, in such a way that the probability measure (\ref{eq:prob_superspins}) is correctly described by the tree-like approximation, and that the BP equations will converge toward a unique fixed-point on a typical large instance.
We give in appendix \ref{app;RS_formalism} the RS equations (\ref{eq:RSeqn}) for the interacting copies of the bi-coloring problem.
These equations can be solved numerically with population dynamics \cite{MezardParisi01}, and provide correct predictions for the typical properties of the measure $\mu_y$ in the regime of small density of constraints $\alpha=M/N$.

\subsubsection{Replica Symmetry Breaking}
\label{subsubsec:setup_RSB}
As the density of constraint $\alpha$ increases, the hypothesis underlying the RS cavity method must break down, and a more sophisticated version of the cavity method can be employed to treat the effect of long loops.
The first non-trivial level is called 1RSB (for one-step Replica Symmetry Breaking), and postulates the existence of a partition of the configuration space into pure states (or clusters) such that the restriction of the measure to one cluster is accurately described within the RS formalism.

The dynamical threshold $\alpha_d$ separates the regime where the RS approximation is valid (for $\alpha<\alpha_d)$ from a region where the 1RSB formalism is needed to correctly describe the typical properties of the measure $\mu_y$.
Technically, the dynamical threshold can be computed within the 1RSB formalism, by deriving and solving the 1RSB equations at Parisi parameter $x=1$ (see appendix \ref{app:1RSB_formalism}, equation (\ref{eq:1RSB_x1})).
These equations always admit a trivial fixed-point: the RS solution given in equation (\ref{eq:RS_fixedpoint_x1}).
In the RS phase, this is the unique solution to the 1RSB equation, correctly describing the case of a single cluster. 
An RSB phase is unveiled by the appearance of a non-trivial solution different from the RS one to the 1RSB equation. 

Depending on the situation, this RS/RSB dynamical transition can occur in a continuous or a discontinuous way. 
In the continuous case, the dynamical threshold $\alpha_d$ can be computed by studying the local instability of the RS solution (see eq.~\ref{eq:RS_fixedpoint_x1}) under a small perturbation toward the space of 1RSB solutions. 
It is called the Kesten-Stigum instability in the context of tree-reconstruction \cite{KeSt66,MoPe03}, or the Almeida-Thouless transition for mean-field spin glasses \cite{AlTh78}. The Kesten-Stigum threshold coincides with the dynamical threshold in the continuous case, and provides an upper bound in the discontinuous case (in which a non-trivial solution to the 1RSB equations appears while the RS solution is still stable).

In order to probe the dynamical phase transition, one can compute the difference between:
\begin{itemize}
	\item the {\it intra-state} overlap $q_1$ (see appendix \ref{app:1RSB_formalism}, equations (\ref{eq:intra_state_q1})) measuring the overlap between two typical configurations sampled from {\it the same} cluster.
	\item and the {\it inter-state} overlap $q_0$ (equation (\ref{eq:inter_state_q0})), measuring the overlap between two typical configurations (in the clustered phase $\alpha>\alpha_d$, two typical configurations are likely to be in two different clusters). In the case of the bi-coloring problem, the inter-state overlap is trivially equal to $0$ due to the spin-flip invariance of the measure (\ref{eq:prob_copies})
\end{itemize}  
While in the RS phase, $q_1-q_0=0$ since the solution space is correctly described by a single cluster, we have $q_1-q_0>0$ in the 1RSB phase (see e.g. Figure \ref{fig:vertical_lines_overlap} the evolution of the intra-state overlap as a function of $\gamma$). If the transition is continuous, the value $q_1-q_0$ grows continuously from 0, otherwise it displays a jump.

\section{Results}

\subsection{Phase diagram}
\label{subsec:phase_diagram}

In this section, we provide a detailed analysis of the effect of the coupling strength $\gamma$ on the dynamical phase transition occurring at $\alpha_d$. 
We restrict our analysis to the case of two copies ($y=2$), and will treat the case of a larger number of copies in a future work.
We also fix the number of variable-per-clause to $k=5$, which is a representative value for the bi-coloring problem for a single copy \cite{CaNaRiZe03} (the cases $k=3$ and $k=4$ being peculiar with a continuous dynamical transition, while it is discontinuous as soon as $k\geq 5$).

\begin{figure}[h!]
    \centering
    \includegraphics[width=\columnwidth]{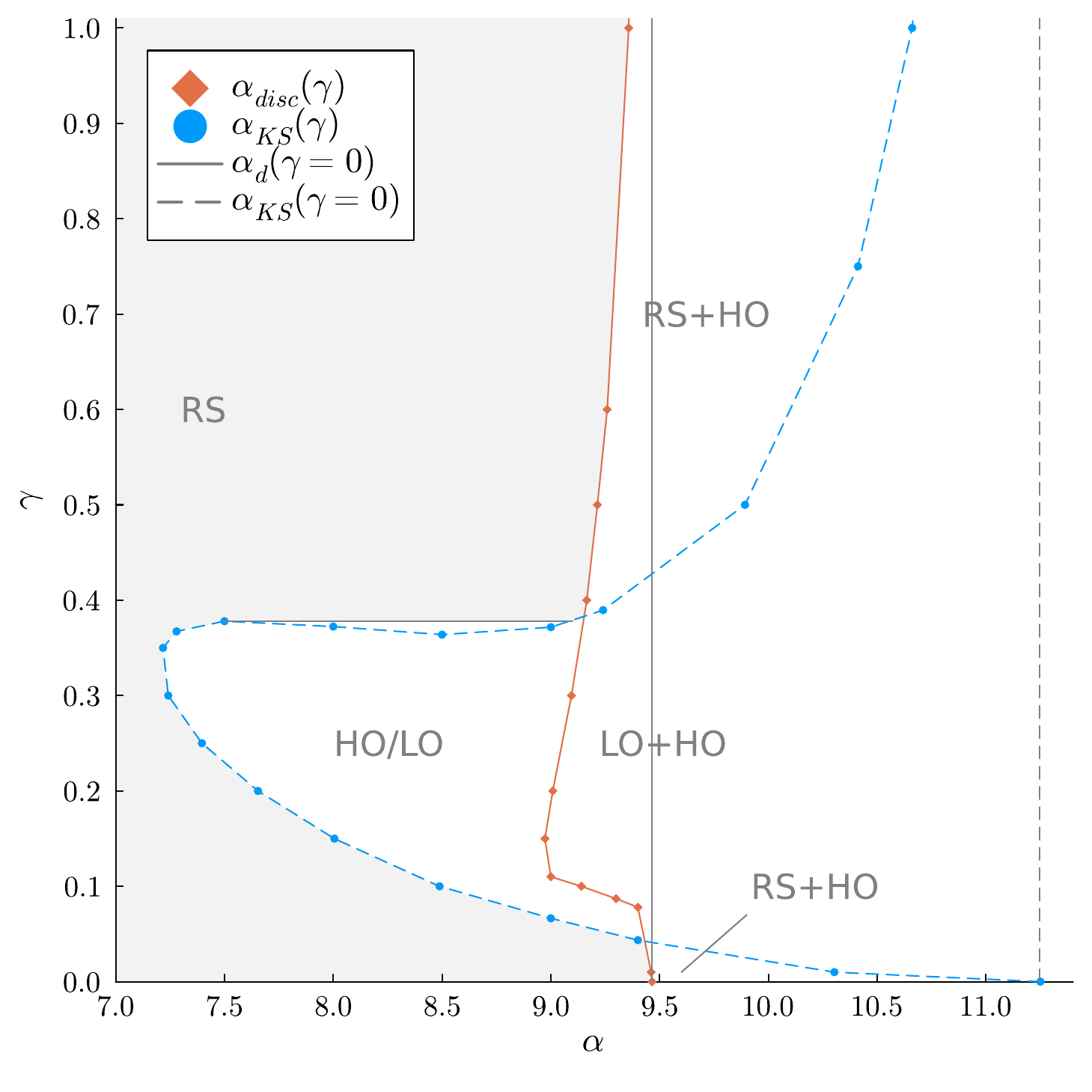}
    \caption{Phase diagram in the plane $(\gamma,\alpha)$ for the replicated bi-coloring problem on random $k$-hypergraphs, with $k=5$ and $y=2$ copies.
    The RS phase is shaded in gray, the RSB phase is in white. 
    The blue circles correspond to the Kesten-Stigum threshold $\alpha_{\rm KS}(\gamma)$.
    The orange squares correspond to the discontinuous appearance of a non-trivial solution to the 1RSB equations at $\alpha_{\rm disc}(\gamma)$. 
    For each value of $\gamma$, the dynamical threshold $\alpha_d(\gamma)=\min\{\alpha_{\rm KS}(\gamma), \alpha_{\rm disc}(\gamma)\}$ separates the RS phase from the RSB phase.
    The vertical solid and dashed lines mark respectively the dynamical and Kesten-Stigum threshold the for the non-interacting case ($\gamma=0$) and in the $\gamma\to\infty$ limit.
    We identify several RSB phases, denoted `LO/HO', `RS+HO', `LO+HO' (see Sec.~\ref{subsubsec:phase_diagram_detailed} for a precise definition).}
    \label{fig:phase_diagram}
\end{figure}

We provide our results in the form of a phase diagram in Fig.~\ref{fig:phase_diagram}, in the $(\alpha,\gamma)$ plane.
In the case $y=2$, the probability distribution (\ref{eq:prob_copies}) satisfy the following symmetry:
\begin{align}
	\label{eq:gamma_symmetry}
	\mu_y(\underline{\sigma}^1,\underline{\sigma}^2;G,\gamma)=\mu_y(\underline{\sigma}^1,-\underline{\sigma}^2;G,-\gamma).
\end{align}
In words, a ferromagnetic system ($\gamma>0$) favoring configurations in which the two copies $\underline{\sigma}^1,\underline{\sigma}^2$ are aligned, is equivalent to the anti-ferromagnetic system ($-\gamma$) favoring the alignment of $\underline{\sigma}^1$ and $-\underline{\sigma}^2$.
This implies that any average quantity computed from this distribution, and in particular the intra-state overlap $q_1$ (\ref{eq:intra_state_q1}), is invariant under the transformation $\gamma\to-\gamma$.
Therefore, the dynamical threshold is such that 
\begin{align}
	\alpha_d(\gamma)=\alpha_d(-\gamma)
\end{align}
and we don't need to explore the region $\gamma<0$.
This symmetry does not hold however for a larger number of copies $y>2$, where both $\gamma>0$ and $\gamma<0$ regions have to be studied.

The RS phase, painted in gray, is on the left of the dynamical threshold $\alpha_d(\gamma)$. 
The blue points correspond to the Kesten-Stigum threshold $\alpha_{\rm KS}(\gamma)$, at which a non-trivial solution to the 1RSB equations emerge continuously from the trivial one upon increasing $\alpha$.
The orange squares mark the discontinuous appearance (at $\alpha_{\rm disc}(\gamma)$) of a non-trivial solution to the 1RSB cavity equations, upon increasing $\alpha$.
Details on the numerical computation of these thresholds are given in appendix \ref{app:subsec_numerical_resolution}.
The dynamical threshold $\alpha_d(\gamma)$, defined as the appearance of a non-trivial solution to the 1RSB equations (\ref{eq:1RSB_x1}), is given as:
\begin{align}
	\alpha_d(\gamma)=\min(\alpha_{\rm KS}(\gamma),\alpha_{\rm disc}(\gamma)).
\end{align}
Note that we recover the threshold values for a single copy ($y=1$) at $\gamma=0$ and in the large $\gamma$ limit: $\alpha_d(y=1)=9.465$ and $\alpha_{\rm KS}(y=1)=11.25$ (see \cite{CaNaRiZe03,DaRa08,GaDaSeZd17,BuRiSe19} for numerical values of the hyper-graph bi-coloring problem' thresholds).

The main observation to extract from this plot is that turning on the coupling between copies has the effect of shrinking the RS phase, for all values of $\gamma\neq0$.
Recalling that large instances can be solved in polynomial time in the RS phase \cite{MoSe06}, this is a negative result: it indicates that the strategy of introducing coupled copies of a CSP might not improve the performance of solving algorithms. This observation is corroborated in the next section \ref{subsec:BP_ongraphs}, where we will see that the BP algorithm fails at converging on finite size instances above the Kesten-Stigum threshold $\alpha_{\rm KS}(\alpha)$. 

In addition, we observe a change in the nature of the phase transition, as the coupling strength $\gamma$ is varied. 
For values of $\gamma$ between $0.04$ and $0.38$, the non-trivial solution to the 1RSB equation appears continuously ($\alpha_{\rm KS}(\gamma)<\alpha_{\rm disc}(\gamma)$), and the dynamical transition is therefore continuous.
Outside of this range, we have $\alpha_{\rm disc}(\gamma)<\alpha_{\rm KS}(\gamma)$, and the dynamical transition is discontinuous. In particular, we recover a discontinuous transition in the non-interacting case (at $\gamma=0$ and in the large $\gamma$ limit), as already observed in \cite{CaNaRiZe03, BuRiSe19} for $k=5$.
On finite-size instances, the effect of a continuous phase transition is more drastic and prevent BP algorithm to converge above $\alpha_d$ (see next section \ref{subsec:BP_ongraphs}). 

\subsubsection{A detailed picture of the phase diagram}
\label{subsubsec:phase_diagram_detailed}
\begin{figure}[h!]
	\centering
	\includegraphics[width=0.4\textwidth]{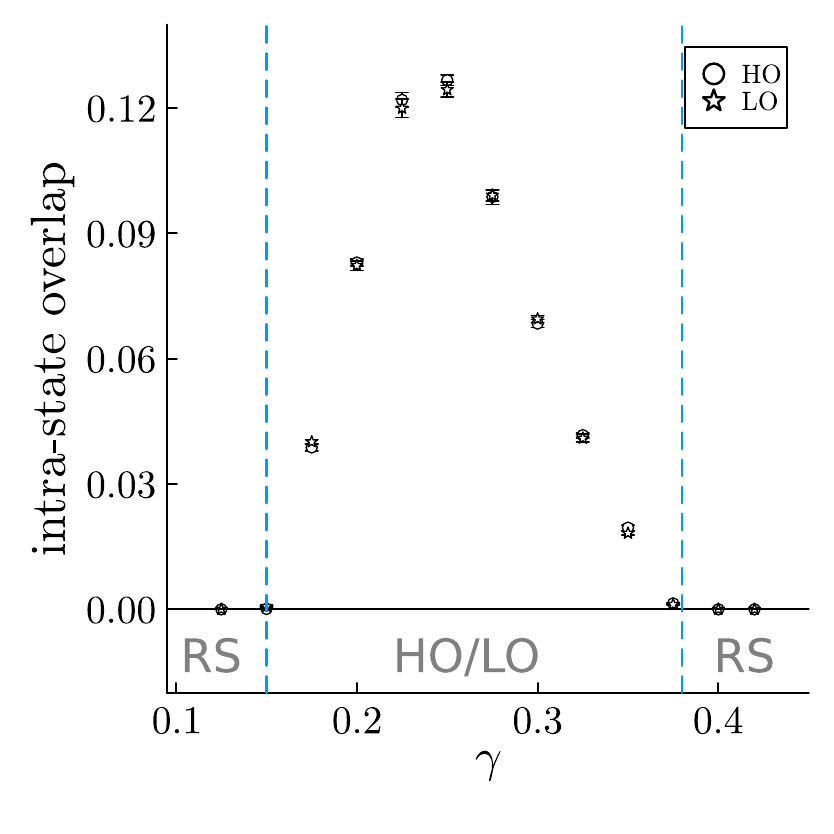}\\
	\includegraphics[width=0.4\textwidth]{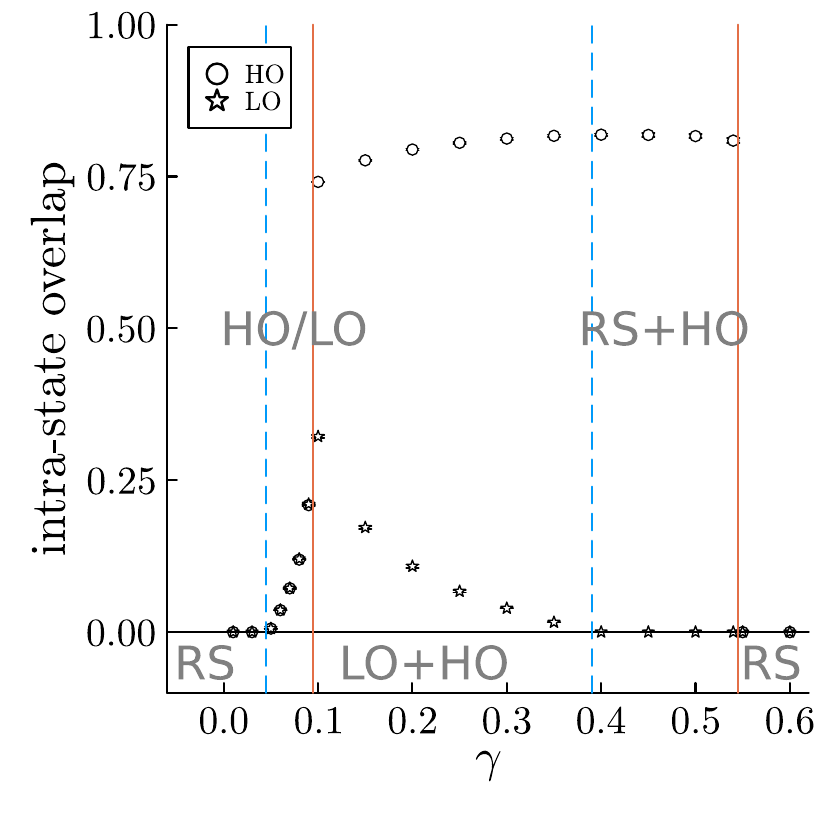}
	\caption{Evolution of the intra-state overlap as a function of the coupling strength $\gamma$, at $y=2, k=5$, for $\alpha=8.0$ (top panel) and $\alpha=9.24$ (bottom panel).
		The blue dashed line marks the Kesten-Stigum threshold, and the orange solid line marks the discontinuous appearance of a non-trivial solution (see Fig.~\ref{fig:phase_diagram}).
	}
	\label{fig:vertical_lines_overlap}
\end{figure}
In this paragraph, we explain the different regions in the phase diagram given in Fig.~\ref{fig:phase_diagram}.
For some values of the parameters $\alpha, \gamma$, there exist (at least) two different solutions to the 1RSB equations at $x=1$ (\ref{eq:1RSB_x1}).
This type of behavior was already observed in \cite{GaDaSeZd17}, \cite{BuRiSe19}, and its consequences for inference problems (and planted CSPs) have been discussed in \cite{RiSeZd19}.
We use two different initial conditions (see appendix \ref{app:subsec_numerical_resolution}) for the iterative resolution of the 1RSB equations:
\begin{itemize}
	\item Starting from a high intra-state overlap $q_1=1$ and denoted `HO' initial condition
	\item Starting with a low intra-state overlap $q_1=0.01$ and denoted `LO' initial condition
\end{itemize}
Depending on the parameters ($\alpha,\gamma$), the two initial conditions can lead to two different solutions, or not. More precisely, the different phases illustrated in Fig.~\ref{fig:phase_diagram} are defined as follows:
\begin{itemize}
	\item RS: both HO and LO initial conditions lead to the trivial RS solution ($q_1=0$);
	\item HO/LO: both HO and LO initial conditions lead to the same non-trivial solution ($q_1>0$);
	\item LO+HO: LO initial condition leads to a non-trivial solution, HO initial condition leads to a different non-trivial solution with a higher overlap;
	\item RS+HO: LO initial condition leads to the trivial RS solution, while HO initial condition leads to a non-trivial solution.
\end{itemize}
The frontiers between these different phases are:
\begin{itemize}
	\item The Kesten-Stigum line: $\alpha_{\rm KS}(\gamma)$, the limit of the stability of the trivial fixed-point under a perturbation toward the space of 1RSB solutions. 
	In the unstable phase, the LO initial condition leads systematically to a non-trivial solution, while in the stable phase it leads to the trivial RS solution.
	\item The discontinuous appearance of a high-overlap (HO) non-trivial solution, at $\alpha_{\rm disc}(\gamma)$, reached from the HO initial condition.
\end{itemize}
In order to get a better understanding of the various phases, we show in Fig.~\ref{fig:vertical_lines_overlap} the evolution of the intra-state overlap as a function of the coupling strength $\gamma$, for two values of $\alpha$.

\subsection{Results on large graph instances}
\label{subsec:BP_ongraphs}
In this section, we compare the results obtained in the large size limit from the cavity method in \ref{subsec:phase_diagram}, with numerical results on finite size instances.

\begin{figure}[h!]
    \centering
    \includegraphics[width=0.4\textwidth]{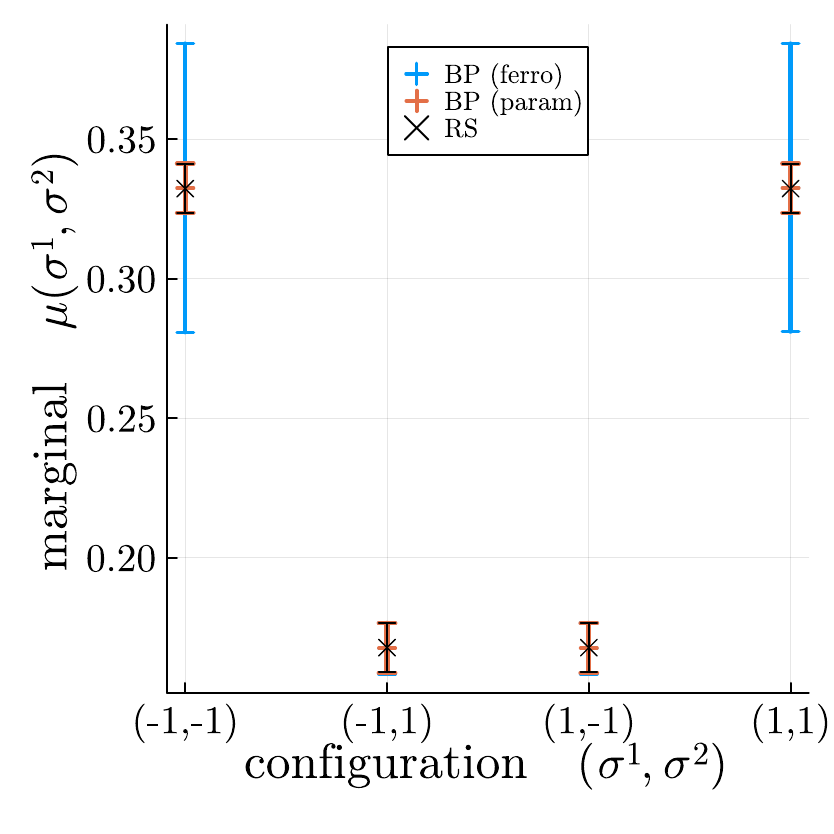} \\
    \includegraphics[width=0.4\textwidth]{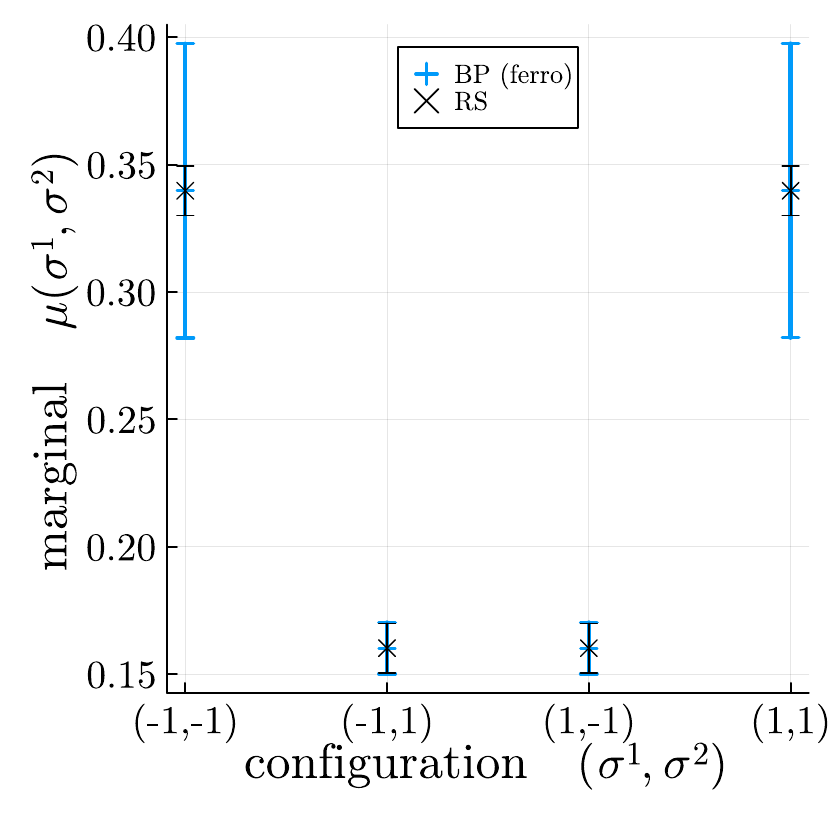}
    \caption{Comparison of the BP marginal averaged over the vertices and over $50$ instances of size $N=10^4$, with BP marginals predicted by the RS formalism. Here $y=2$, $k=5$, $\gamma=0.2$.
    Top panel: for $\alpha=7.5$, a {\it paramagnetic} BP solution is found on a large fraction (0.84) of instances, and the corresponding paramagnetic BP marginals (in orange) are comparable with the RS prediction (in black). 
    A {\it ferromagnetic} BP solution is found only on a small fraction (0.16) of instances, and their BP marginals (in blue) are quantitatively different from RS. 
    Bottom panel: for $\alpha=7.7$, BP marginals are ferromagnetic for all instances (in blue), and quantitatively different from the RS predictions (in black)}
    \label{fig:BP_marginals}
\end{figure}

We start by comparing the BP marginals $\mu_i(\sigma_i^1,\sigma_i^2)$ found on finite size instances, with the  RS solution in the large size limit.
In Fig.~\ref{fig:BP_marginals} we report this comparison at $\gamma=0.2$, for two values of $\alpha$, one below the dynamical threshold $\alpha_d(\gamma=0.2)=7.654$ (top panel) and one above $\alpha_d$ (bottom panel).
We compute the BP marginals in Eq.~(\ref{eq:BPmarg}) for each value of $X_i=(\sigma_i^1,\sigma_i^2)\in\{(+1,+1),(+1,-1),(-1,+1),(-1,-1)\}$, and average over the vertices of the graph, and over $50$ graph instances of size $N=10^4$.
The error bars reported in Fig.~\ref{fig:BP_marginals} represent the standard deviation of $\mu_i(X_i)$ over the population for the RS solution (in black) and over the instances and the vertices of the graph for the BP solutions (in orange and blue).

For any number of copies $y$, a simple criterion for distinguishing between a ferromagnetic (polarized) and a paramagnetic BP fixed-point is given by the following quantity
\begin{align}
\begin{aligned}
{\rm FP} &= \frac{1}{N}\sum_{i=1}^N {\rm FP}_i \quad \text{with}\\
{\rm FP}_i&=\frac{1}{y}\sum_{s=1}^y \left( \mu_i^s(1)^2 + \mu_i^s(-1)^2\right)\;,
\end{aligned}
\end{align}
where $\mu_i^s(\sigma)=\mathbb{P}[\sigma_i^s=\sigma]$ is the single variable marginal on site $i$ in replica $s$.
In a paramagnetic solution, we have $\mu_i^s=1/2$ for each vertex $i\in \{1,\dots,N\}$ and for each copy $s\in\{1,\dots,y\}$, and therefore ${\rm FP}=0.5$, while ${\rm FP}>0.5$ for a ferromagnetic solution.

At $\alpha=7.5$, we observe two types of BP fixed-points, ferromagnetic (in blue) and paramagnetic (in orange).
At $\alpha=7.7$, a ferromagnetic BP fixed-point is always found on all instances.
Furthermore, there is a good compatibility between the {\it paramagnetic} BP marginal found on finite size-instances at $\alpha=7.5$ (Fig.~\ref{fig:BP_marginals}, top panel, in orange), and the RS prediction in the large size limit (in black).
The small error on the BP marginals is a finite size effect that should go to zero for larger graphs.

Instead, the {\it ferromagnetic} BP marginals are quantitatively different from the RS prediction.
Although, for each value of $X\in\{(+1,+1),(+1,-1),(-1,+1),(-1,-1)\}$ the average value of $\mu(X)$ is remarkably close to the RS prediction, the fluctuations of $\mu(1,1)$ and $\mu(-1,-1)$ over the graph sites and over the instances are much larger.
In practice the population representing the RS solution is much more homogeneous that the BP marginals on given large graphs.

\begin{figure}[t]
    \centering
    \includegraphics[width=0.4\textwidth]{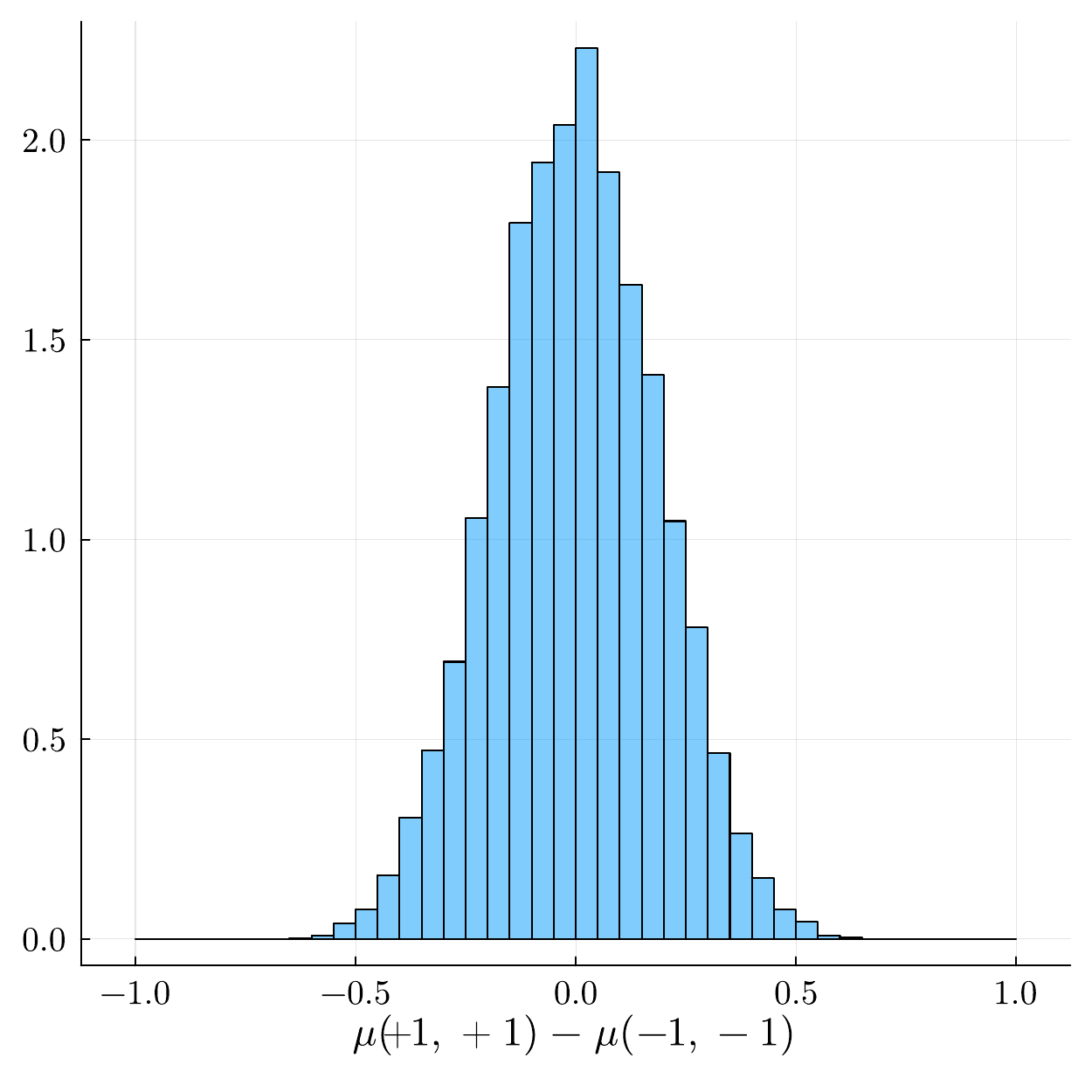}
    \caption{Histogram of the quantity $\mu_i(+1,+1)-\mu_i(-1,-1)$ for each site of a given instance of size $N=10^4$. Here $y=2$, $k=5$, $\gamma=0.2$, $\alpha=7.7$.}
    \label{fig:histoDiff}
\end{figure}

Given that the largest fluctuations are in values of $\mu(1,1)$ and $\mu(-1,-1)$, and are highly anti-correlated (i.e., the sum of the two values remains almost constant, close to 0.68), we report in Fig.~\ref{fig:histoDiff} the histogram of the difference $\mu(1,1)-\mu(-1,-1)$.
We see such a difference span the whole possible range, and gives evidence that on a given large graph the BP solution predicts the presence of strongly polarized vertices, that is vertices whose marginal probability is concentrated on a single value $(1,1)$ or $(-1,-1)$.

The presence of these polarized vertices is a clear sign of an RSB phase transition, as already found for spin glass models at low temperatures \cite{perrupato2022ising}. 
Indeed, due to the global spin-flip invariance of the probability measure (\ref{eq:prob_superspins}), we have $\mu_y(\underline{X})=\mu_y(-\underline{X})$, and also the marginal probability on any vertex $i\in V$ is also invariant under spin-flip, $\mu_i(X_i) = \mu_i(-X_i)$.
As a consequence, on each vertex $i$, the marginal probability $\mu_i$ is not polarized, and one should always observe a paramagnetic solution with $\text{FP}=0.5$.

Therefore, a {\it ferromagnetic} BP fixed-point with $\text{FP}>0.5$ is not describing correctly the full probability distribution (\ref{eq:prob_superspins}).
This ferromagnetic solution rather describe a restricted part of the probability distribution, that is a {\it cluster}), in which some vertices are polarized toward one specific value (while all vertices should be not polarized once averaged over the clusters).
Finding a ferromagnetic BP fixed-point on some instances therefore could be a signal of the RSB phase transition occurring at the dynamical threshold $\alpha_d=7.654$, and the appearance of glassy states even below this value could be understood as a finite-size effect (on some graph instances, a polarized solution might naturally appear).

We want to stress that standard BP with a single replica never finds ferromagnetic fixed points, neither in the clustered phase, nor in the paramagnetic phase. These ferromagnetic FPs are only found by the replicated BP: it could be possible that the replicated measure enlarges the basin of attraction of the glassy states with larger entropy that become thus accessible for the replicated BP. 

\begin{figure}[t]
    \centering
    \includegraphics[width=0.4\textwidth]{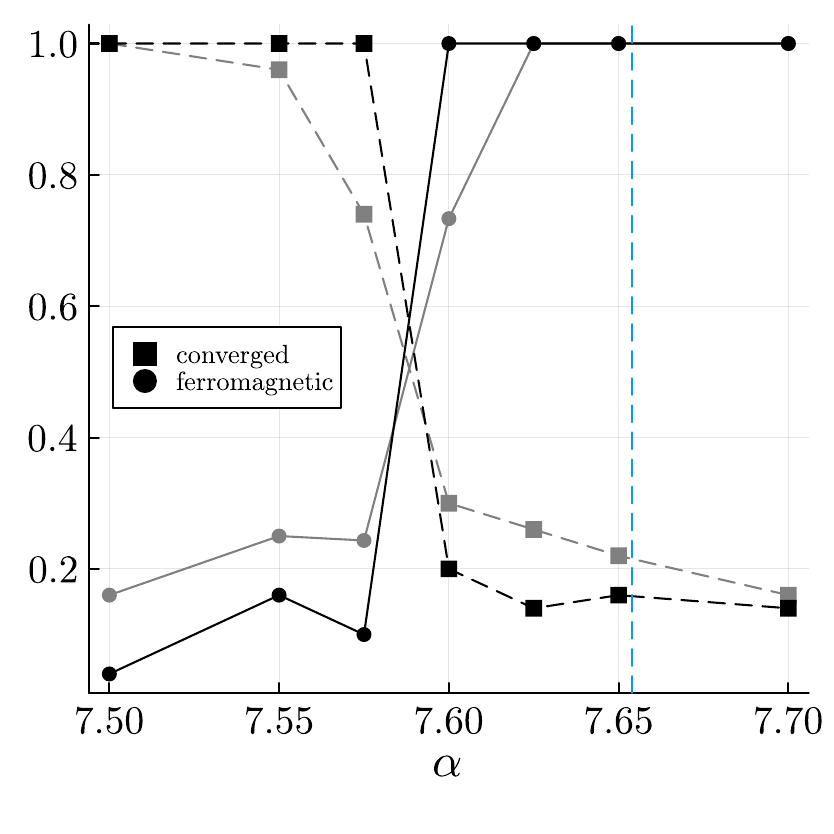} \\
    \includegraphics[width=0.4\textwidth]{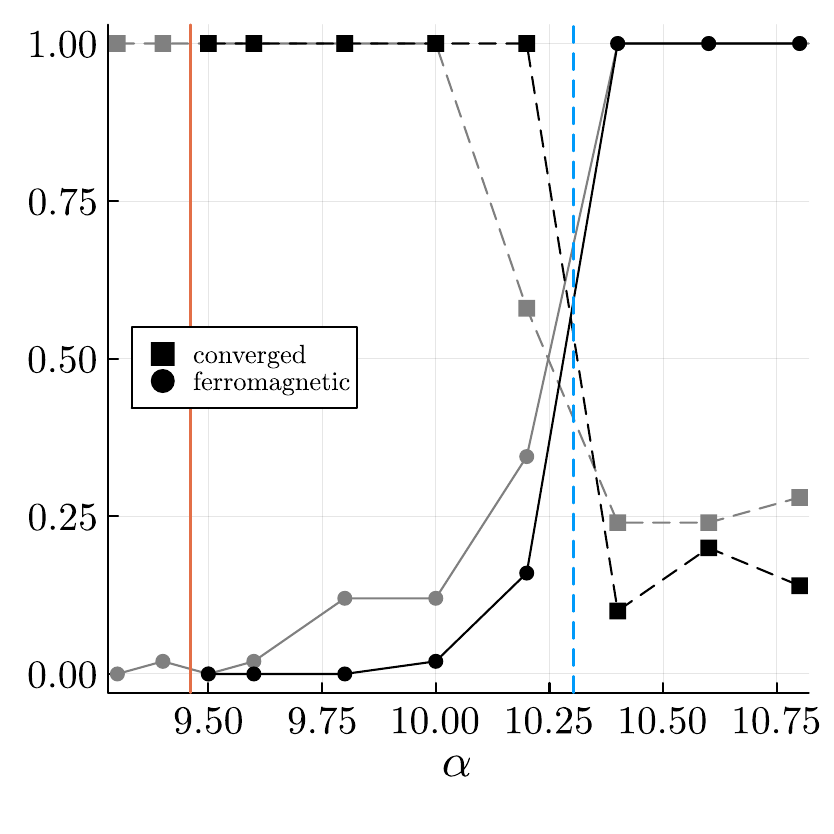}
    \caption{Solutions found by BP on finite-size instances, for $y=2$, $k=5$, at $\gamma=0.2$ (top panel) and $\gamma=0.01$ (bottom panel), for two graph sizes: $N=10^4$ (gray points), and $N=10^5$ (black points). 
    We display the fraction of instances, over $50$ runs, for which BP converged (squares), and, among converged instances, the fraction of ferromagnetic BP fixed-points (circles).
    The vertical lines correspond to the thresholds computed in Sec.~\ref{subsec:phase_diagram}: $\alpha_{\rm KS}(\gamma)$ (blue dashed line), $\alpha_{\rm disc}(\gamma)$ (orange solid line). 
    The last threshold is not displayed on the top panel, as it occurs at larger values $\alpha_{\rm disc}(\gamma=0.2)=9.01$.}
    \label{fig:BP_conv}
\end{figure}

Fig.~\ref{fig:BP_conv} illustrates the appearance of the ferromagnetic solution as $\alpha$ approaches the dynamical threshold $\alpha_d(\gamma)$, for two values of $\gamma$: 
\begin{enumerate}[(a)]
	\item $\gamma=0.2$: at which the dynamical phase transition is continuous
	(see phase diagram Fig.~\ref{fig:phase_diagram} and explanations in Sec.~\ref{subsec:phase_diagram}).
	\item $\gamma=0.01$: at which the dynamical phase transition is discontinuous.
\end{enumerate}

We considered two graph sizes ($N=10^4, 10^5$), and run BP on $50$ random graphs for each size. 
We show in Fig.~\ref{fig:BP_conv} the fraction of graphs for which BP iterations converged towards a fixed-point (squares), and among the converged runs, the fraction of ferromagnetic BP fixed-points (circles).
In general, at the BP algorithmic transition, the fraction of converged BP runs drastically decreases and the few converged runs achieve a ferromagnetic fixed point.

In the case of a continuous dynamical transition at $\gamma=0.2$ (top panel), we observe the BP algorithmic transition to take place slightly below the dynamical threshold $\alpha_d(\gamma=0.2)=\alpha_{\rm KS}(\gamma=0.2)=7.65$, the small difference being possibly a finite size effect.

In the case of a discontinuous transition at $\gamma=0.01$, the dynamical threshold coincides with the discontinuous appearance of a non-trivial RSB solution at $\alpha_d(\gamma=0.01)=\alpha_{\rm disc}(\gamma=0.01)=9.46$, but it does not affect the behavior of BP.
The BP algorithmic behavior is again ruled by the Kesten-Stigum threshold: only when the paramagnetic fixed point becomes locally unstable, BP fails to converge or converges to a ferromagnetic fixed point.

We recall that the Kesten-Stigum threshold decreases as soon as the coupling between copies is turned on with $\gamma\neq 0$ (see the phase diagram in Fig.~\ref{fig:phase_diagram}, blue dashed line).
Introducing a coupling between copies has therefore the effect of shrinking the range of $\alpha$ for which BP can converge on finite size instances.

The above results may be interpreted as an indication that the coupling strategy can not increase the performance of message-passing algorithms in solving an instance of CSP.
However, we have to remind that the BP algorithm by itself does not output a solution, but only a set of marginal probability distributions (\ref{eq:BPmarg}).
In order to obtain a solution to a given instance of CSP, the BP algorithm should be followed by a decimation procedure.
An analysis of the convergence of BP during the decimation procedure can be made following the steps of \cite{MoRiSe07, Ricci-Tersenghi_2009}, and we plan to do it in a future work.

\section{Discussion and Perspectives}

We have introduced a model of $y$ interacting copies of a constraint satisfaction problem, the bi-coloring problem on $k$-hyper-graphs, and studied the effect of the coupling strength between copies on the dynamical phase transition.
We focused on the case of two copies ($y=2$), and leave for future work the analysis of this model for a larger (but finite) number of copies, as well as an analytical study in the large $y$ limit.

We showed that turning on the coupling strength has the effect of shrinking the Replica Symmetric phase, where algorithms exist that can sample solutions in polynomial time.
This result is quite surprising, because it was conjectured that a re-weighting strategy favoring solutions living in dense clusters (i.e., with a large local entropy \cite{BaInLuSaZe15_long}) should enhance algorithmic performances \cite{BaBo16}. This conjecture was further confirmed by numerical investigations mainly in inference problems, with a planted solution \cite{AnRi23, CaDeSe24}.
However, the conjecture was not tested at large in optimization problems, as in the present case.
Only in Ref.~\cite{angelini2019monte} it was shown that for the largest independent set optimization problem, introducing coupled replicas does not improve standard algorithms such as Simulated Annealing.

An interesting feature that we have observed is the change of the nature of the clustering phase transition, from discontinuous to continuous (in the range $\gamma\in[0.04,0.38]$).
This may have important algorithmic consequences.
Indeed, in a continuous phase transition, we expect that approximating the RSB solution slightly beyond the phase transition is easier than finding the RSB solution that appears in a discontinuous phase transition.
It would be very interesting to check whether a Monte Carlo based algorithm, like SA, could work around the continuous phase transition, while we expect it to fail dramatically at a discontinuous phase transition.

We also studied the effect of the re-weighting strategy induced by the coupling on the behavior of the BP algorithm on finite-size instances, and find that the convergence of BP is strongly affected by the continuous transition occurring at the Kesten-Stigum threshold.
These findings call for a better understanding of the behavior of algorithms at the clustering transition.
A promising direction could be to perform an analysis of the BP plus the decimation procedure, following the formalism developed in \cite{MoRiSe07, Ricci-Tersenghi_2009}, adapting it to the case of $y$ interacting copies.

Another interesting direction will be to consider the associated planted problem (or inference problem), as it was shown \cite{AnRi23} that an algorithmic strategy based on replicated interacting copies of the problem (replicated SA) could outperform classical SA and reach the Bayes optimal threshold.

In addition, we also plan to extend our results when finite temperature is added to the problem. This should be particularly useful to explain the behaviour of algorithms, such as Simulated Annealing, that do not work directly at zero temperature.

\acknowledgments
We thank Guilhem Semerjian for insightful discussions.
The research has received financial support from the ``National Centre for HPC, Big Data and Quantum Computing - HPC'', Project CN\_00000013, CUP B83C22002940006, NRP Mission 4 Component 2 Investment 1.5, Funded by the European Union - NextGenerationEU, and from the 2021 FIS (Fondo Italiano per la Scienza) funding scheme (FIS783 - SMaC - Statistical Mechanics and Complexity) and the PRIN funding scheme (2022LMHTET - Complexity, disorder and
fluctuations: spin glass physics and beyond), both from the Italian MUR (Ministry of University and Research).

\bibliography{biblio.bib}

\begin{thebibliography}{10}

\bibitem{MonassonZecchina99b}
R.~Monasson, R.~Zecchina, S.~Kirkpatrick, B.~Selman, and L.~Troyansky.
\newblock 2+p-SAT: Relation of typical-case complexity to the nature of the
  phase transition.
\newblock {\em Random Structures and Algorithms}, {\bfseries 15}, 414 (1999).

\bibitem{BiroliMonasson00}
G.~Biroli, R.~Monasson, and M.~Weigt.
\newblock A variational description of the ground state structure in random
  satisfiability problems.
\newblock {\em Eur. Phys. J. B}, {\bfseries 14}, 551 (2000).

\bibitem{MezardParisi02}
M.~M\'ezard, G.~Parisi, and R.~Zecchina.
\newblock Analytic and Algorithmic Solution of Random Satisfiability Problems.
\newblock {\em Science}, {\bfseries 297}, 812--815 (2002).

\bibitem{MertensMezard06}
S.~Mertens, M.~M\'ezard, and R.~Zecchina.
\newblock Threshold values of random K-SAT from the cavity method.
\newblock {\em Random Struct. Algorithms}, {\bfseries 28}(3), 340--373 (2006).

\bibitem{krzakala2007gibbs}
F.~Krzakala, A.~Montanari, F.~Ricci-Tersenghi, G.~Semerjian, and L.~Zdeborova.
\newblock Gibbs states and the set of solutions of random constraint
  satisfaction problems.
\newblock {\em Proceedings of the National Academy of Sciences}, {\bfseries
  104}(25), 10318--10323 (2007).

\bibitem{mezard2009information}
M.~Mezard and A.~Montanari.
\newblock {\em Information, physics, and computation}.
\newblock Oxford University Press, 2009.

\bibitem{AchlioptasRicci06}
D.~Achlioptas and F.~Ricci-Tersenghi.
\newblock On the solution-space geometry of random constraint satisfaction
  problems.
\newblock In {\em Proc. of 38th STOC}, pages 130--139, New York, NY, USA, 2006.
  ACM.

\bibitem{AchlioptasCoja-Oghlan08}
D.~Achlioptas and A.~Coja-Oghlan.
\newblock Algorithmic Barriers from Phase Transitions.
\newblock In {\em 2008 49th Annual IEEE Symposium on Foundations of Computer
  Science}, pages 793--802, 2008.

\bibitem{molloy_col_freezing}
M.~Molloy.
\newblock The freezing threshold for k-colourings of a random graph.
\newblock In {\em Proceedings of the 44th symposium on Theory of Computing},
  page 921. ACM, 2012.

\bibitem{ding2014proof}
J.~Ding, A.~Sly, and N.~Sun.
\newblock Proof of the Satisfiability Conjecture for Large K.
\newblock In {\em Proceedings of the Forty-seventh Annual ACM Symposium on
  Theory of Computing}, STOC '15, pages 59--68, 2015.

\bibitem{Gamarnik_21}
D.~Gamarnik.
\newblock The overlap gap property: A topological barrier to optimizing over
  random structures.
\newblock {\em Proceedings of the National Academy of Sciences}, {\bfseries
  118}(41), e2108492118 (2021).

\bibitem{GaSu17}
D.~Gamarnik and M.~Sudan.
\newblock Performance of Sequential Local Algorithms for the Random
  NAE-\$K\$-SAT Problem.
\newblock {\em SIAM Journal on Computing}, {\bfseries 46}(2), 590--619 (2017).

\bibitem{CoHaHe17}
A.~Coja-Oghlan, A.~Haqshenas, and S.~Hetterich.
\newblock Walksat Stalls Well Below Satisfiability.
\newblock {\em SIAM Journal on Discrete Mathematics}, {\bfseries 31}(2),
  1160--1173 (2017).

\bibitem{BrHu22}
G.~Bresler and B.~Huang.
\newblock The Algorithmic Phase Transition of Random k-SAT for Low Degree
  Polynomials.
\newblock In {\em 2021 IEEE 62nd Annual Symposium on Foundations of Computer
  Science (FOCS)}, pages 298--309, 2022.

\bibitem{gamarnik2025turing}
D.~Gamarnik.
\newblock Turing in the shadows of Nobel and Abel: an algorithmic story behind
  two recent prizes.
\newblock {\em arXiv preprint arXiv:2501.15312},  (2025).

\bibitem{KirkpatrickGelatt83}
S.~Kirkpatrick, C.~D. {Gelatt Jr.}, and M.~P. Vecchi.
\newblock Optimization by Simulated Annealing.
\newblock {\em Science}, {\bfseries 220}, 671--680 (1983).

\bibitem{AnRi23}
M.~C. Angelini and F.~Ricci-Tersenghi.
\newblock Limits and Performances of Algorithms Based on Simulated Annealing in
  Solving Sparse Hard Inference Problems.
\newblock {\em Phys. Rev. X}, {\bfseries 13}, 021011 (2023).

\bibitem{MoRiSe07}
A.~Montanari, F.~Ricci-Tersenghi, and G.~Semerjian.
\newblock Solving Constraint Satisfaction Problems through Belief
  Propagation-guided decimation.
\newblock {\em 45th Annual Allerton Conference on Communication, Control, and
  Computing 2007}, {\bfseries 1} (2007).

\bibitem{Ricci-Tersenghi_2009}
F.~Ricci-Tersenghi and G.~Semerjian.
\newblock On the cavity method for decimated random constraint satisfaction
  problems and the analysis of belief propagation guided decimation algorithms.
\newblock {\em Journal of Statistical Mechanics: Theory and Experiment},
  {\bfseries 2009}(09), P09001 (2009).

\bibitem{angelini2025algorithmic}
M.~Angelini, M.~Avila-Gonz{\'a}lez, F.~D'Amico, D.~Machado, R.~Mulet, and
  F.~Ricci-Tersenghi.
\newblock Algorithmic thresholds in combinatorial optimization depend on the
  time scaling.
\newblock {\em arXiv preprint arXiv:2504.11174},  (2025).

\bibitem{MoSe06}
A.~Montanari and G.~Semerjian.
\newblock Rigorous Inequalities Between Length and Time Scales in Glassy
  Systems.
\newblock {\em Journal of Statistical Physics}, {\bfseries 125}(1), 23--54
  (2006).

\bibitem{BuRiSe19}
L.~Budzynski, F.~Ricci-Tersenghi, and G.~Semerjian.
\newblock Biased landscapes for random constraint satisfaction problems.
\newblock {\em Journal of Statistical Mechanics: Theory and Experiment},
  {\bfseries 2019}(2), 023302 (2019).

\bibitem{BrDaSeZd16}
A.~Braunstein, L.~Dall’Asta, G.~Semerjian, and L.~Zdeborová.
\newblock The large deviations of the whitening process in random constraint
  satisfaction problems.
\newblock {\em Journal of Statistical Mechanics: Theory and Experiment},
  {\bfseries 2016}(5), 053401 (2016).

\bibitem{BaInLuSaZe15_long}
C.~Baldassi, A.~Ingrosso, C.~Lucibello, L.~Saglietti, and R.~Zecchina.
\newblock Local entropy as a measure for sampling solutions in constraint
  satisfaction problems.
\newblock {\em Journal of Statistical Mechanics: Theory and Experiment},
  {\bfseries 2016}(2), 023301 (2016).

\bibitem{BaBo16}
C.~Baldassi, C.~Borgs, J.~T. Chayes, A.~Ingrosso, C.~Lucibello, L.~Saglietti,
  and R.~Zecchina.
\newblock Unreasonable effectiveness of learning neural networks: From
  accessible states and robust ensembles to basic algorithmic schemes.
\newblock {\em Proceedings of the National Academy of Sciences}, {\bfseries
  113}(48), E7655--E7662 (2016).

\bibitem{MaSeSeZa18}
T.~Maimbourg, M.~Sellitto, G.~Semerjian, and F.~Zamponi.
\newblock Generating dense packings of hard spheres by soft interaction design.
\newblock {\em SciPost Phys.}, {\bfseries 4}, 39 (2018).

\bibitem{BuSe20}
L.~Budzynski and G.~Semerjian.
\newblock Biased measures for random constraint satisfaction problems: larger
  interaction range and asymptotic expansion.
\newblock {\em Journal of Statistical Mechanics: Theory and Experiment},
  {\bfseries 2020}(10), 103406 (2020).

\bibitem{ZhZh20}
H.~Zhao and H.-J. Zhou.
\newblock Maximally flexible solutions of a random $K$-satisfiability formula.
\newblock {\em Phys. Rev. E}, {\bfseries 102}, 012301 (2020).

\bibitem{CaDeSe24}
G.~Catania, A.~Decelle, and B.~Seoane.
\newblock Copycat perceptron: Smashing barriers through collective learning.
\newblock {\em Phys. Rev. E}, {\bfseries 109}, 065313 (2024).

\bibitem{CaNaRiZe03}
T.~Castellani, V.~Napolano, F.~Ricci-Tersenghi, and R.~Zecchina.
\newblock Bicolouring random hypergraphs.
\newblock {\em Journal of Physics A: Mathematical and General}, {\bfseries
  36}(43), 11037 (2003).

\bibitem{MezardParisi01}
M.~M\'ezard and G.~Parisi.
\newblock The Bethe lattice spin glass revisited.
\newblock {\em Eur. Phys. J. B}, {\bfseries 20}, 217 (2001).

\bibitem{MezardParisi03}
M.~M\'ezard and G.~Parisi.
\newblock The cavity method at zero temperature.
\newblock {\em J. Stat. Phys.}, {\bfseries 111}, 1--34 (2003).

\bibitem{KeSt66}
H.~Kesten and B.~P. Stigum.
\newblock {Additional Limit Theorems for Indecomposable Multidimensional
  Galton-Watson Processes}.
\newblock {\em The Annals of Mathematical Statistics}, {\bfseries 37}(6), 1463
  -- 1481 (1966).

\bibitem{MoPe03}
E.~Mossel and Y.~Peres.
\newblock {Information flow on trees}.
\newblock {\em The Annals of Applied Probability}, {\bfseries 13}(3), 817 --
  844 (2003).

\bibitem{AlTh78}
J.~R.~L. de~Almeida and D.~J. Thouless.
\newblock Stability of the Sherrington-Kirkpatrick solution of a spin glass
  model.
\newblock {\em Journal of Physics A: Mathematical and General}, {\bfseries
  11}(5), 983 (1978).

\bibitem{DaRa08}
L.~Dall'Asta, A.~Ramezanpour, and R.~Zecchina.
\newblock Entropy landscape and non-Gibbs solutions in constraint satisfaction
  problems.
\newblock {\em Phys. Rev. E}, {\bfseries 77}, 031118 (2008).

\bibitem{GaDaSeZd17}
M.~Gabrié, V.~Dani, G.~Semerjian, and L.~Zdeborová.
\newblock Phase transitions in the q-coloring of random hypergraphs.
\newblock {\em Journal of Physics A: Mathematical and Theoretical}, {\bfseries
  50}(50), 505002 (2017).

\bibitem{RiSeZd19}
F.~Ricci-Tersenghi, G.~Semerjian, and L.~Zdeborov\'a.
\newblock Typology of phase transitions in Bayesian inference problems.
\newblock {\em Phys. Rev. E}, {\bfseries 99}, 042109 (2019).

\bibitem{perrupato2022ising}
G.~Perrupato, M.~C. Angelini, G.~Parisi, F.~Ricci-Tersenghi, and T.~Rizzo.
\newblock Ising spin glass on random graphs at zero temperature: Not all spins
  are glassy in the glassy phase.
\newblock {\em Physical Review B}, {\bfseries 106}(17), 174202 (2022).

\bibitem{angelini2019monte}
M.~C. Angelini and F.~Ricci-Tersenghi.
\newblock Monte Carlo algorithms are very effective in finding the largest
  independent set in sparse random graphs.
\newblock {\em Physical Review E}, {\bfseries 100}(1), 013302 (2019).

\bibitem{montanari2008clusters}
A.~Montanari, F.~Ricci-Tersenghi, and G.~Semerjian.
\newblock Clusters of solutions and replica symmetry breaking in random
  k-satisfiability.
\newblock {\em Journal of Statistical Mechanics: Theory and Experiment},
  {\bfseries 2008}(04), P04004 (2008).

\end{thebibliography}

\appendix

\section{Cavity method under the Replica Symmetric ansatz}
\label{app;RS_formalism}

In the RS formalism, one assumes that the effect of long loops in a sample of the hyper-graph ensemble is negligible, and that the BP equations (\ref{eq:BPeqn}) admit a unique fixed-point that correctly describes the probability distribution (\ref{eq:prob_superspins}).
We consider a uniformly chosen directed edge $i\to a$ on a random hyper-graph $G$, and let $\mathcal{P}^{\rm RS}(\eta)$ be the probability distribution of the fixed-point variable-to-factor BP message $\eta_{i\to a}$ thus obtained. 
Similarly, we define the distribution $\widehat{\mathcal{P}}^{\rm RS}(\widehat{\eta})$ of the factor-to-variable BP message $\widehat{\eta}_{a\to i}$. Then, under the RS hypothesis, the incoming messages on a given variable node $i\in V$ (resp. a factor node $a\in E$) are i.i.d. with the law $\widehat{\mathcal{P}}^{\rm RS}$ (resp. $\mathcal{P}^{\rm RS}$). This implies that the laws $\mathcal{P}^{\rm RS}$ and $\widehat{\mathcal{P}}^{\rm RS}$ must obey the following equations:
\begin{align}
\label{eq:RSeqn}
\begin{aligned}
\mathcal{P}^{\rm RS}(\eta)&=\sum_{d=1}^\infty r_d\int \prod_{a=1}^d{\rm d}\widehat{\mathcal{P}}^{\rm RS}(\widehat{\eta}_a)\delta[\eta-f^{\rm BP}(\widehat{\eta}_1,\dots,\widehat{\eta}_d)]\\
\widehat{\mathcal{P}}^{\rm RS}(\widehat{\eta})&=\int\prod_{i=1}^{k-1}{\rm d}\mathcal{P}^{\rm RS}(\eta_i)\delta[\widehat{\eta}-\widehat{f}^{\rm BP}(\eta_1,\dots,\eta_{k-1})]
\end{aligned}
\end{align}
where $\eta=f^{\rm BP}(\widehat{\eta}_1,\dots,\widehat{\eta}_d)$ and $\widehat{\eta}=\widehat{f}^{\rm BP}(\eta_1,\dots,\eta_{k-1})$ are shorthand notations for the BP equations (\ref{eq:BPeqn}), and where $r_d$ is the residual degree distribution:
\begin{align}
r_d=\frac{(d+1)p_d}{\sum_d (d+1)p_d}
\end{align}
with $p_d$ the degree distribution. For the Erd\"os R\'enyi ensemble, the degree distribution follows a Poisson law of mean $\alpha k$, and the residual distribution therefore follows the same law.

\subsection{RS Overlap}
We define the following component-wise overlap between two super-spin configurations $\underline{X}, \underline{X}'$ for each $s\in\{1,\dots,y\}$:
\begin{align}
	\label{eq:overlap_s}
O_s(\underline{X}, \underline{X}') =\frac{1}{N}\sum_{i=1}^N \sigma_i^s\sigma_i^{'s}
\end{align}
where $\underline{X}=(X_1,\dots,X_N)$, and $X_i=(\sigma_i^1,\dots,\sigma_i^y)\in\{-1,+1\}^y$.
The averaged overlap over the super-spin probability distribution (\ref{eq:prob_superspins}) writes:
\begin{align}
	\label{eq:averaged_overlap_s}
	\begin{aligned}
\langle O_s(\underline{X}, \underline{X}')\rangle_{\mu_y} &= \sum_{\underline{X}}\sum_{\underline{X}'}\mu_y(\underline{X})\mu_y(\underline{X}')O_s(\underline{X}, \underline{X}')\\
&= \frac{1}{N}\sum_{i=1}^N\left(\sum_{X_i}\mu_i(X_i)\sigma_i^s\right)^2
	\end{aligned}
\end{align}
with $\mu_i$ the marginal probability distribution of $X_i$.

In order to compute the overlap averaged over the random ensemble of hyper-graph instances, we need to average over the distribution of BP marginals $\mu_i$.
In the RS formalism, we obtain (with $X=(\sigma^1,\dots,\sigma^y)\in\{-1,1\}^y)$:
\begin{align}
q_s^{\rm RS} = \int{\rm d}\mathcal{Q}^{\rm RS}(\mu)\left(\sum_X\mu(X)\sigma^s\right)^2
\end{align}
with $\mathcal{Q}^{\rm RS}$ probability distribution for the BP marginal $\mu_i$, for a uniformly chosen node $i$ on a random hyper-graph $G$, satisfying the equation:
\begin{align}
\mathcal{Q}^{\rm RS}(\mu)&=\sum_d p_d\int \prod_{a=1}^d{\rm d}\widehat{\mathcal{P}}^{\rm RS}(\widehat{\eta}_a)\delta[\mu-g^{\rm BP}(\widehat{\eta}_1,\dots,\widehat{\eta}_d)]
\end{align}
with $\mu=g^{\rm BP}(\widehat{\eta}_1,\dots,\widehat{\eta}_d)$ a shorthand notation for equation (\ref{eq:BPmarg}).

\subsection{Invariances of the RS solution}

Note that the probability distribution (\ref{eq:prob_superspins}) is invariant under a global spin-flip symmetry.
The RS prediction for the overlap $q_s^{RS}$ is therefore trivially equal to $0.5$. 
It is however a useful sanity check to compute this quantity in order to verify the correctness of the RS solution found numerically with population dynamics.

Similarly, the probability distribution (\ref{eq:prob_superspins}) is invariant is invariant under a permutation of the $y$ copies $\underline{X}^1,\dots,\underline{X}^y$:
\begin{align}
\mu_y(\underline{X}^1,\dots,\underline{X}^y) = \mu_y(\underline{X}^{\pi(1)},\dots,\underline{X}^{\pi(y)}) \qquad \forall \pi\in\mathcal{S}(y)
\end{align}
As a result, the marginal probabilities $\mu_i(X_i)$ are also invariant under a permutation of its components. This is verified for instance in Fig.~\ref{fig:BP_marginals}, and it is also a good sanity-check for the RS solution found numerically.\\

\section{One-step Replica Symmetry Breaking cavity method}
\label{app:1RSB_formalism}

\subsection{1RSB cavity equations}
Under the 1RSB hypothesis, one assumes that the probability distribution (\ref{eq:prob_superspins}) is partitioned into clusters (or pure states):
\begin{align}
	\mu_y(\underline{X}) = \sum_{\gamma}p(\gamma)\mu_{y,\gamma}(\underline{X})
\end{align}
with $p(\gamma)$ the distribution over the clusters. The restriction of the distribution $\mu_y$ to one cluster can be described by the RS formalism, i.e. can be described by a fixed-point of the BP equations.

We define $P_{i\to a}$ (resp. $\widehat{P}_{a\to i}$) as the probability law of the message $\eta_{i\to a}^\gamma$ (resp $\widehat{\eta}_{a\to i}^\gamma$), for a cluster $\gamma$ being chosen randomly with probability $p(\gamma)$.  
Then the 1RSB messages $P_{i\to a},\widehat{P}_{a\to i}$ satisfy the following self-consistent equations:
\begin{widetext}
\begin{align}
\label{eq:1RSB_1graph}
\begin{aligned}
P_{i\to a}(\eta_{i\to a}) &=\frac{1}{Z^{\rm 1RSB}_{i\to a}}\int\prod_{b\in\partial i\setminus a}{\rm d}\widehat{P}_{b\to i}(\widehat{\eta}_{b\to i})\delta\left(\eta_{i\to a}-f^{\rm BP}(\{\widehat{\eta}_{b\to i}\}_{b\in\partial i\setminus a})\right)z_{i\to a}(\{\widehat{\eta}_{b\to i}\}_{b\in\partial i\setminus a})^x \\
\widehat{P}_{a\to i}(\widehat{\eta}_{a\to i}) &=\frac{1}{\widehat{Z}_{a\to i}^{\rm 1RSB}}\int\prod_{j\in\partial a\setminus i} {\rm d}P_{j\to a}(\eta_{j\to a})\delta\left(\widehat{\eta}_{a\to i} - \widehat{f}^{\rm BP}(\{\eta_{j\to a}\}_{j\in\partial a\setminus i})\right)\widehat{z}_{a\to i}(\{\{\eta_{j\to a}\}_{j\in\partial a\setminus i}\})^x
\end{aligned}
\end{align}
\end{widetext}
where $x$ is the Parisi parameter, allowing to weight
differently the various clusters according to their size.
In the above equation, $Z^{\rm 1RSB}_{i\to a}$ and $\widehat{Z}^{\rm 1RSB}_{a\to i}$ are normalization factors for $P_{i\to a}$ and $\widehat{P}_{a\to i}$. The terms $z_{i\to a}(\{\widehat{\eta}_{b\to i\setminus a}\}_{b\in\partial i})$ and $\widehat{z}_{a\to i}(\{\eta_{j\to a}\}_{j\in\partial a\setminus i})$ are the normalization factors in equations (\ref{eq:BPeqn}).

In order to average over the disorder, one introduces the probability distributions over the 1RSB messages: $\mathcal{P}^{\rm 1RSB}(P)$ and $\widehat{\mathcal{P}}^{\rm 1RSB}(\widehat{P})$. They obey the 1RSB equations (similar to the RS equations (\ref{eq:RSeqn})):
\begin{align}
\label{eq:1RSBeqn}
\begin{aligned}
\mathcal{P}^{\rm RSB}(P) &= \sum_dr_d\int\prod_{a=1}^d{\rm d}\widehat{\mathcal{P}}^{\rm 1RSB}(\widehat{P}_a)\delta(P-F(\widehat{P}_1\dots \widehat{P}_d)) \\
\widehat{\mathcal{P}}^{\rm RSB}(\widehat{P}) &= \int\prod_{i=1}^{k-1}{\rm d}\mathcal{P}^{\rm 1RSB}(P_i)\delta(Q-\widehat{F}(P_1,\dots,P_{k-1}))
\end{aligned}
\end{align}
where $P=F(\widehat{P}_1,\dots,\widehat{P}_d)$ and $\widehat{P}=\widehat{F}(P_1,\dots,P_{k-1})$ are shorthand notations for the equations (\ref{eq:1RSB_1graph}).

\subsection{RS trivial fixed-point}
On a given graph, in the Replica Symmetric phase, there is only one fixed point to the BP equation (\ref{eq:BPeqn}), that we denote $\{\bar{\eta}_{i\to a}, \bar{\widehat{\eta}}_{a\to i}\}_{i\in V, a\in\partial i}$.
The solution to the equations (\ref{eq:1RSB_1graph}) is therefore a trivial Dirac delta:
\begin{align}
\begin{aligned}
P_{i\to a}(\eta_{i\to a}) &= \delta(\eta_{i\to a},\bar{\eta}_{i\to a})\\
\widehat{P}_{i\to a}(\widehat{\eta}_{a\to i}) &= \delta(\widehat{\eta}_{a\to i},\bar{\widehat{\eta}}_{a\to i})
\end{aligned}
\end{align}
Once averaged over the disorder, we can see that the 1RSB equations (\ref{eq:1RSBeqn}) always admit the RS trivial fixed-point:
\begin{align}
\label{eq:RS_fixedpoint}
\begin{aligned}
\mathcal{P}^*(P) &= \int {\rm d}\mathcal{P}^{\rm RS}(\bar{\eta})\delta[P(\eta)-\delta(\eta,\bar{\eta})] \\
\widehat{\mathcal{P}}^*(\widehat{P}) &= \int {\rm d}\widehat{\mathcal{P}}^{\rm RS}(\bar{\widehat{\eta}})\delta[\widehat{P}(\widehat{\eta})-\delta(\widehat{\eta},\bar{\widehat{\eta}})]
\end{aligned}
\end{align}
For small values of the density of constraints $\alpha$, this trivial solution is the only one, and the predictions given by the RS and 1RSB cavity method coincide: we are in the RS phase.
Increasing $\alpha$, non-trivial solutions of (\ref{eq:1RSBeqn}) can appear. 
The clustering threshold $\alpha_d(y, \gamma)$ is defined as the smallest value of $\alpha$ for which the 1RSB equations at $x=1$ admit a non-trivial solution.

\subsection{Simplifications of the 1RSB equations at Parisi parameter $x=1$}
The complete 1RSB equations (\ref{eq:1RSBeqn}) can be simplified for the special value of $x=1$.
As explained in \cite{montanari2008clusters}, the first step is to note that the normalization factor $Z_{i\to a}^{\rm 1RSB}$ (resp.$\widehat{Z}_{a\to i}^{\rm 1RSB}$) depend only on the distributions $\{\widehat{P}_{b\to i}\}_{b\in\partial i \setminus a}$ (resp. on $\{P_{j\to a}\}_{j\in\partial a \setminus i}$) through their mean value.
One defines $\bar{\eta}[P]$, $\bar{\widehat{\eta}}[\widehat{P}]$ as the averages: 
\begin{align}
	\label{eq:bar_eta}
\begin{aligned}
\bar{\eta}[P](X) &= \int{\rm d}P(\eta)\eta(X) \\
\bar{\widehat{\eta}}[\widehat{P}](X) &= \int{\rm d}\widehat{P}(\widehat{\eta})\widehat{\eta}(X)
\end{aligned}
\end{align}
Then, one can check that $Z_{i\to a}^{\rm 1RSB}$ depends on the distributions $\widehat{P}_1,\dots,\widehat{P}_d$ only through the averages $\bar{\widehat{\eta}}[\widehat{P}_1],\dots,\bar{\widehat{\eta}}[\widehat{P}_d]$ (and similarly for $\widehat{Z}_{a\to i}^{\rm 1RSB}$):
\begin{align}
\begin{aligned}
Z_{i\to a}^{\rm 1RSB}(\widehat{P}_1,\dots,\widehat{P}_d) &= z_{i\to a}(\bar{\widehat{\eta}}[\widehat{P}_1],\dots,\bar{\widehat{\eta}}[\widehat{P}_d])\\
\widehat{Z}_{a\to i}^{\rm 1RSB}(P_1,\dots ,P_{k-1}) &= \widehat{z}_{a\to i}(\bar{\eta}[P_1],\dots,\bar{\eta}[P_{k-1}])
\end{aligned}
\end{align}
Then, one can check that the random variables $\bar{\eta}[P],\bar{\widehat{\eta}}[\widehat{P}]$ obtained by drawing $P$ (resp $\widehat{P}$) from $\mathcal{P}^{\rm 1RSB}(P)$ (resp $\mathcal{\widehat{P}}^{\rm 1RSB}(\widehat{P})$) satisfy the RS equations (\ref{eq:RSeqn}), and therefore are distributed according to the RS distributions $\mathcal{P}^{\rm RS}, \mathcal{\widehat{P}}^{\rm RS}$.
One defines the conditional averages:
\begin{align}
	\label{eq:unconditional_averages}
\begin{aligned}
\bar{P}(\eta|\bar{\eta}) &= \frac{1}{\mathcal{P}^{\rm RS}(\bar{\eta})}\int{\rm d}\mathcal{P}^{\rm 1RSB}(P)P(\eta)\delta(\bar{\eta},\bar{\eta}[P])\\
\bar{\widehat{P}}(\widehat{\eta}|\bar{\widehat{\eta}}) &= \frac{1}{\mathcal{\widehat{P}}^{\rm RS}(\bar{\widehat{\eta}})}\int{\rm d}\mathcal{\widehat{P}}^{\rm 1RSB}(\widehat{P})\widehat{P}(\widehat{\eta})\delta(\bar{\widehat{\eta}},\bar{\widehat{\eta}}[\widehat{P}])
\end{aligned}
\end{align}
We can get closed equations for these distributions:
\begin{widetext}
\begin{align}
	\begin{aligned}
	\bar{P}(\eta|\bar{\eta})\mathcal{P}^{\rm RS}(\bar{\eta}) &=\sum_dr_d \int\prod_{a=1}^{d}{\rm d}\mathcal{\widehat{P}}^{\rm RS}(\bar{\widehat{\eta}}_a)\delta[\bar{\eta}-f^{\rm BP}(\bar{\widehat{\eta}}_1,\dots \bar{\widehat{\eta}}_d)]\int \prod_{a=1}^d{\rm d}\bar{\widehat{P}}(\widehat{\eta}_a|\bar{\widehat{\eta}}_a)\delta[\eta-f^{\rm BP}(\widehat{\eta}_1,\dots \widehat{\eta}_d)]\frac{z_{i\to a}(\widehat{\eta}_1,\dots \widehat{\eta}_d))}{z_{i\to a}(\bar{\widehat{\eta}}_1,\dots \bar{\widehat{\eta}}_d))}\\
	\bar{\widehat{P}}(\widehat{\eta}|\bar{\widehat{\eta}})\mathcal{\widehat{P}}^{\rm RS}(\bar{\widehat{\eta}}) &= \int\prod_{i=1}^{k-1}{\rm d}\mathcal{P}^{\rm RS}(\bar{\eta}_i)\delta[\bar{\widehat{\eta}}-\widehat{f}^{\rm BP}(\bar{\eta}_1,\dots \bar{\eta}_{k-1})]\int \prod_{i=1}^{k-1}{\rm d}\bar{P}(\eta_i|\bar{\eta}_i)\delta[\widehat{\eta}-\widehat{f}^{\rm BP}(\eta_1,\dots \eta_{k-1})]\frac{\widehat{z}_{a\to i}(\eta_1,\dots \eta_{k-1}))}{\widehat{z}_{a\to i}(\bar{\eta}_1,\dots \bar{\eta}_{k-1}))}
	\end{aligned}
\end{align}
\end{widetext}
These equations are simpler than equations (\ref{eq:1RSBeqn}): the distribution $\bar{P}(\eta|\bar{\eta})\mathcal{P}^{\rm RS}(\bar{\eta})$ can be seen as a joint distribution over $\eta, \bar{\eta}$, and can be represented by a population of couples ($\{\eta_i, \bar{\eta}_i\}_{i=1}^\mathcal{N}$). However, the factor $\frac{z_{i\to a}(\widehat{\eta}_1,\dots \widehat{\eta}_d))}{z_{i\to a}(\bar{\widehat{\eta}}_1,\dots \bar{\widehat{\eta}}_d))}$ is still hard to represent with a population. 
To get rid of it, we can define the following distributions (following the steps of \cite{montanari2008clusters}):
\begin{align}
	\label{eq:1RSB_conditional_distribs}
	\begin{aligned}
	P_X(\eta|\bar{\eta}) &= \frac{\eta(X)}{\bar{\eta}(X)}\bar{P}(\eta|\bar{\eta})\\
	\widehat{P}_X(\widehat{\eta}|\bar{\widehat{\eta}}) &= \frac{\widehat{\eta}(X)}{\bar{\widehat{\eta}}(X)}\bar{\widehat{P}}(\widehat{\eta}|\bar{\widehat{\eta}})
	\end{aligned}
\end{align}
One can check that these distributions now satisfy the following self-consistent equations:
\begin{widetext}
\begin{align}
\label{eq:1RSB_x1}
\begin{aligned}
P_{X}(\eta|\bar{\eta})\mathcal{P}^{\rm RS}(\bar{\eta}) &= \sum_dr_d\int\prod_{a=1}^d{\rm d}\mathcal{\widehat{P}}^{\rm RS}(\bar{\widehat{\eta}}_a)\delta[\bar{\eta}-f^{\rm BP}(\bar{\widehat{\eta}}_1,\dots\bar{\widehat{\eta}}_d)]\int\prod_{a=1}^d{\rm d}\widehat{P}_X(\widehat{\eta}_a|\bar{\widehat{\eta}}_a)\delta[\eta-f^{\rm BP}(\widehat{\eta}_1,\dots\widehat{\eta}_d)]\\
\widehat{P}_{X}(\widehat{\eta}|\bar{\widehat{\eta}})\mathcal{\widehat{P}}^{\rm RS}(\bar{\widehat{\eta}}) &= \int\prod_{i=1}^{k-1}{\rm d}\mathcal{P}^{\rm RS}(\bar{\eta}_i)\delta[\bar{\widehat{\eta}}-\widehat{f}^{\rm BP}(\bar{\eta}_1,\dots\bar{\eta}_{k-1})]\sum_{\{X_i\}_{i=1}^k}\nu(\{X_i\}_{i=1}^{k-1}|X,\{\bar{\eta}_i\}_{i=1}^{k-1})\int\prod_{i=1}^{k-1}{\rm d}P_{X_i}(\eta_i|\bar{\eta}_i)\\
&\qquad\qquad\qquad\qquad\qquad\qquad\qquad\qquad\qquad\qquad\qquad\qquad\qquad\times\delta[\widehat{\eta}-\widehat{f}^{\rm BP}(\eta_1,\dots\eta_{k-1})]
\end{aligned}
\end{align}
with the probability distribution:
\begin{align}
	\label{eq:distrib_nu}
\nu(\{X_i\}_{i=1}^{k-1}|X,\{\bar{\eta}_i\}_{i=1}^{k-1})=\frac{\Omega(X,\{X_i\}_{i=1}^{k-1})\prod_{i=1}^{k-1}\bar{\eta}_i(X_i)}{\sum_{\{X_i\}_{i=1}^{k-1}}\Omega(X,\{X_i\}_{i=1}^{k-1})\prod_{i=1}^{k-1}\bar{\eta}_i(X_i)}\,.
\end{align}
\end{widetext}
Equations (\ref{eq:1RSB_x1}) can be solved with population dynamics. One employs a population of tuples $\{\bar{\eta}_i,\{\eta_i^{X}\}_{X\in\{+1,-1\}^y}\}_{i=1}^\mathcal{N}s$ and $\{\bar{\widehat{\eta}}_i,\{\widehat{\eta}_i^{X}\}_{X\in\{+1,-1\}^y}\}_{i=1}^\mathcal{N}s$. As each tuple contains $2^y+1$ elements, this representation can become too heavy to be implemented numerically with a large value of $y$. 
We will see in the following (see section \ref{subsec:simplifs_symmetries}) how to decrease the number of elements to be stored by using the symmetries of the model.

\subsubsection{RS trivial fixed-point}
The RS trivial fixed-point (\ref{eq:RS_fixedpoint}) can be written in terms of the distributions (\ref{eq:1RSB_conditional_distribs}):
\begin{align}
\label{eq:RS_fixedpoint_x1}
\begin{aligned}
P_X(\eta|\bar{\eta})\mathcal{P}^{\rm RS}(\bar{\eta})=\delta(\eta,\bar{\eta})\mathcal{P}^{\rm RS}(\bar{\eta})\\
\widehat{P}_X(\widehat{\eta}|\bar{\widehat{\eta}})\mathcal{\widehat{P}}^{\rm RS}(\bar{\widehat{\eta}})=\delta(\widehat{\eta},\bar{\widehat{\eta}})\mathcal{\widehat{P}}^{\rm RS}(\bar{\widehat{\eta}})\,.
\end{aligned}
\end{align}

\subsubsection{Further simplifications using symmetries}
\label{subsec:simplifs_symmetries}
Let $\pi:\{1,\dots,y\}\to\{1,\dots,y\}$ be a permutation of the super-spin components. Let $X=(\sigma^1,\dots,\sigma^y)\in\{+1,-1\}^y$ be a super-spin vector. We will use the notation
$\pi(X)=(\sigma^{\pi(1)},\dots,\sigma^{\pi(y)})$. 
Furthermore, let $\eta^\pi$ be the BP message $\eta$ after permutation of the components:
\begin{align}
	\eta^\pi(X) = \eta(\pi(X))\, .
\end{align}
Starting from equations (\ref{eq:unconditional_averages}) one can check that the probability distribution $\bar{P}(\eta|\bar{\eta})$ satisfies the following invariance:
\begin{align}
	\bar{P}(\eta|\bar{\eta}) = \bar{P}(\eta^\pi|\bar{\eta}^\pi)
\end{align}
(and similarly for $\bar{\widehat{P}}(\widehat{\eta}|\bar{\widehat{\eta}})$).

Using the definitions (\ref{eq:1RSB_conditional_distribs}) for the distributions $P_X$ and $\widehat{P}_X$, this invariance translates into the relations:
\begin{align}
	\label{eq:invariance_P_X}
	\begin{aligned}	
	P_X(\eta|\bar{\eta})&=P_{\pi^{-1}(X)}(\eta^\pi|\bar{\eta}^\pi)\\
	\widehat{P}_X(\widehat{\eta}|\bar{\widehat{\eta}})&=\widehat{P}_{\pi^{-1}(X)}(\widehat{\eta}^\pi|\bar{\widehat{\eta}}^\pi)
	\end{aligned}
\end{align}
One can use this relation to decrease the number of distributions involved in the self-consistent equations (\ref{eq:1RSB_x1}).

For each $n\in\{0,1,\dots,y\}$, let 
\begin{align}
\label{eq:def_U_n}
U_n = (\underbrace{1,\dots,1}_{n},\underbrace{-1,\dots,-1}_{y-n})
\end{align}
be the configuration with the $n$ first components being $1$, the remaining components being $-1$.
For each $X\in\{+1,-1\}^y$, define $\pi_X$ such that:
\begin{align}
	\pi_X(X) = U_{n(X)}
\end{align}
where $n(X)=|\{s\in\{1,\dots,y\}: \sigma(X)^s=1\}|$ is the number of components equal to $1$ in $X$. In other words, $\pi_X$ is the permutation of the components such that, when applied to the vector $X$, it places all the $1$-components first.

Define the $y+1$ distributions, for each $n\in\{0,1,\dots,y\}$:
\begin{align}
	P_n=P_{U_n} \qquad \text{(and similarly:}\quad \widehat{P}_n=\widehat{P}_{U_n}\text{)}
\end{align}
Then, the invariance relation (\ref{eq:invariance_P_X}) applied to $\pi=\pi_X$ gives:
\begin{align}
	P_X(\eta|\bar{\eta}) = P_{n(X)}\left(\eta^{\left(\pi_X\right)^{-1}}|\bar{\eta}^{\left(\pi_X\right)^{-1}}\right)
\end{align}
Inserting this relation into the second equation of (\ref{eq:1RSB_x1}), this allows us to write a set of self consistent equations only between the $2(y+2)$ distributions $\{\widehat{P}_n\}_{n\in\{0,\dots,y\}},\{\widehat{P}_n\}_{n\in\{0,\dots,y\}},\mathcal{P}^{\rm RS}$, and $\mathcal{\widehat{P}}^{\rm RS}$:
\begin{widetext}
	\begin{align}
		\label{eq:1RSB_x1_sym}
		\begin{aligned}
			P_n(\eta|\bar{\eta})\mathcal{P}^{\rm RS}(\bar{\eta}) &= \sum_dr_d\int\prod_{a=1}^d{\rm d}\mathcal{\widehat{P}}^{\rm RS}(\bar{\widehat{\eta}}_a)\delta[\bar{\eta}-f^{\rm BP}(\bar{\widehat{\eta}}_1,\dots\bar{\widehat{\eta}}_d)]\int\prod_{a=1}^d{\rm d}\widehat{P}_n(\widehat{\eta}_a|\bar{\widehat{\eta}}_a)\delta[\eta-f^{\rm BP}(\widehat{\eta}_1,\dots\widehat{\eta}_d)]\\
			\widehat{P}_n(\widehat{\eta}|\bar{\widehat{\eta}})\mathcal{\widehat{P}}^{\rm RS}(\bar{\widehat{\eta}}) &= \int\prod_{i=1}^{k-1}{\rm d}\mathcal{P}^{\rm RS}(\bar{\eta}_i)\delta[\bar{\widehat{\eta}}-\widehat{f}^{\rm BP}(\bar{\eta}_1,\dots\bar{\eta}_{k-1})]\sum_{\{X_i\}_{i=1}^k}\nu(\{X_i\}_{i=1}^{k-1}|X,\{\bar{\eta}_i\}_{i=1}^{k-1})\int\prod_{i=1}^{k-1}{\rm d}P_{n(X_i)}(\eta_i|\bar{\eta}_i)\\
			&\qquad\qquad\qquad\qquad\qquad\qquad\qquad\qquad\qquad\qquad\qquad\qquad\qquad\times\delta[\widehat{\eta}-\widehat{f}^{\rm BP}(\eta_1^{\pi_{X_1}},\dots\eta^{\pi_{X_{k-1}}}_{k-1})]
		\end{aligned}
	\end{align}
\end{widetext}

We solved these equations iteratively with population dynamics, see appendix \ref{app:subsec_numerical_resolution} for more details on the numerical resolution. 

\subsubsection{Inter-state and intra-state overlap}
The inter-state overlap $q_0$ and intra-state overlap $q_1$, defined in section \ref{subsubsec:setup_RSB} can be written in terms of the 1RSB distributions. 
Starting from the definition of the component-wise overlap (\ref{eq:overlap_s}) and its average (\ref{eq:averaged_overlap_s}) over the super-spin probability distribution (\ref{eq:prob_superspins}), we obtain:
\begin{itemize}
	\item the (component-wise) {\it inter-state overlap} between two configurations sampled independently from (\ref{eq:prob_superspins}):
	\begin{align}
		\label{eq:inter_state_q0}
		q_0^s = \int {\rm d}\mathcal{P}^{\rm 1RSB}(P)\left[\left(\int{\rm d}P(\eta)\sum_{X\in\{\pm 1\}^y}\sigma^s\eta(X)\right)^2\right]
	\end{align}
	with $X=(\sigma^1,\dots,\sigma^y)\in\{+1,-1\}^y$.
	\item the (component-wise) {\it intra-state overlap} between two configurations sampled from the same cluster:
	\begin{align}
		\label{eq:intra_state_q1}
		q_1^s = \int {\rm d} \mathcal{P}^{\rm 1RSB}(P)\left[\int{\rm d}P(\eta)\left(\sum_{X\in\{\pm 1\}^y}\sigma^s\eta(X)\right)^2\right]
	\end{align}	
\end{itemize}
The intra-state and inter-state overlap $q_0$ and $q_1$ are simply
$q_0= \sum_{s=1}^yq_0^s$, $q_1= \sum_{s=1}^yq_1^s$. In particular, $q_1$ is the quantity plotted in Fig.~\ref{fig:vertical_lines_overlap}.

One can check that the inter-state overlap is trivially equal to the RS overlap $q_s^{\rm RS}$. Using the definition of $\bar{\eta}[P]$ (\ref{eq:bar_eta}), and recalling that $\bar{\eta}[P]$ satisfies the RS equations (\ref{eq:RSeqn}), therefore is distributed according to $\mathcal{P}^{\rm RS}$, one obtains:
\begin{align}
	\begin{aligned}
	q_0^s &= \int {\rm d}\mathcal{P}^{\rm 1RSB}(P) \left(\sum_{X\in\{\pm 1\}^y}\sigma^s\bar{\eta}[P](X)\right)^2\\
	&= \int	{\rm d}\mathcal{P}^{\rm RS}(\eta)\left(\sum_{X\in\{\pm 1\}^y}\sigma^s\eta(X)\right)^2 = q_s^{\rm RS}.
	\end{aligned}
\end{align}

Furthermore, one can express the intra-state overlap in terms of the distribution $P_X$:
\begin{align}
	\begin{aligned}
	q_1^s &= \int {\rm d} \mathcal{P}^{\rm RS}(\bar{\eta})\left[\sum_X \sigma^s\bar{\eta}(X)\int {\rm d}P_X(\eta|\bar{\eta})\left(\sum_{X'}\eta(X')\sigma'^s\right)\right] \\
	&= \int{\rm d}\mathcal{P}^{\rm RS}(\bar{\eta})\left[\sum_X \sigma^s\bar{\eta}(X)\int {\rm d}P_{n(X)}(\eta|\bar{\eta})\left(\sum_{X'}\eta^{\pi_X}(X')\sigma'^s\right)\right].
	\end{aligned}
\end{align}
With $X=(\sigma^1,\dots,\sigma^y)$ and $X'=(\sigma'^1,\dots,\sigma'^y)$. In the second equation, we used the invariance relations (\ref{eq:invariance_P_X}) to write the intra-state overlap only in terms of the distributions $\{P_n\}_{n\in\{0,\dots,y\}}$.

\subsection{Numerical resolution of the 1RSB equations}
\label{app:subsec_numerical_resolution}
The numerical resolution of the 1RSB equations (\ref{eq:1RSB_x1_sym}) can be done with population dynamics.
A first possibility is to represent both the distributions $\{\mathcal{P}^{\rm}, P_0, \dots, P_y\}$ and $\{\mathcal{\widehat{P}}^{\rm}, \widehat{P}_0, \dots, \widehat{P}_y\}$, with two population of tuples: $\{(\bar{\eta}^{(i)},\eta_0^{(i)},\dots,\eta_y^{(i)}): i\in\{1,\dots,\mathcal{N}\}\}$ and $\{(\bar{\widehat{\eta}}^{(i)},\widehat{\eta}_0^{(i)},\dots,\widehat{\eta}_y^{(i)}): i\in\{1,\dots,\mathcal{N}\}\}$.
However, an iterative solution using this implementation suffers from oscillations that prevent convergence to a fixed point. 
We therefore preferred to use a single population of elements $\{(\bar{\eta}^{(i)},\eta_0^{(i)},\dots,\eta_y^{(i)}): i\in\{1,\dots,\mathcal{N}\}\}$.
This implementation amounts to solve the set of self-consistent equations on the distributions $\{\mathcal{P}^{\rm}, P_0, \dots, P_y\}$ only (that can be obtained by plugging the second equation of (\ref{eq:1RSB_x1_sym}) into the first one).

At each iteration, one constructs a new population of elements by repeating for $i\in\{1,\dots,\mathcal{N}\}$, independently, the following steps: 
\begin{itemize}
	\item draw a random number $d$ from the residual distribution $r_d$ (in the case of ER graphs, it is a Poisson law of parameter $\alpha k$).
	\item draw $d(k-1)$ indices $\{j_{i,a}\}_{i\in\{1,\dots, k-1\}, a\in\{1,\dots,d\}}$ i.i.d. from $\{1,\dots,\mathcal{N}\}$
	\item Compute the new message:
	\begin{align}
	\bar{\eta}^{(i)}=\tilde{f}^{\rm BP}(\{\bar{\eta}^{(j_{i,a})}\}_{i\in\{1,\dots, k-1\}, a\in\{1,\dots,d\}})
	\end{align}
	with $\tilde{f}^{\rm BP}$ a shorthand notation for the BP equation on the $\eta$ messages only (obtained by plugging the second equation of (\ref{eq:BPeqn}) in the first one). This can be done efficiently with convolutions, see next section \ref{app:convolutions} for more details.
	\item For each $n\in\{0,\dots,y\}$:
	\begin{itemize}
		\item Sample, for each $a\in\{1,\dots,d\}$ independently, the super-spin configurations: $\{X_{i,a}\}_{i\in\{1,\dots,k-1\}}$ from the distribution (cf. equation (\ref{eq:distrib_nu}))
	 	$$
	 	\nu(\cdot|U_n,\{\bar{\eta}_{j_{i,a}}\}_{i\in\{1,\dots,k-1\}})
	 	$$ 
	 	with $U_n = (\underbrace{1,\dots,1}_{n},\underbrace{-1,\dots,-1}_{y-n})$ be the configuration with the $n$ first components being $1$, the remaining components being $-1$ (see section \ref{app:sampling}. for an efficient implementation).
	 	\item Apply the permutation to each incoming message:
	 	$$
	 	\eta'^{(j_{i,a})}_{n(X_{i,a})} = \left(\eta^{(j_{i,a})}_{n(X_{i,a})}\right)^{\pi_{X_{i,a}}}
	 	$$
		\item Compute the new message:
	\begin{align}
		\eta_n^{(i)}=\tilde{f}^{\rm BP}(\{\eta'^{(j_{i,a})}_{n(X_{i,a})}\}_{i\in\{1,\dots, k-1\}, a\in\{1,\dots,d\}})
	\end{align}
	\item Fill the population with the new element $\{(\bar{\eta}^{(i)},\eta_0^{(i)},\dots,\eta_y^{(i)}): i\in\{1,\dots,\mathcal{N}\}\}$.
	\end{itemize}
\end{itemize}
Iterating these steps many times, one converges to a fixed point solution of (\ref{eq:1RSB_x1_sym}).

\subsubsection{Convolutions}
\label{app:convolutions}
In this sub-section, one explains how to compute a new message $\widehat{\eta}=\widehat{f}^{\rm BP}(\{\eta_i\}_{i\in\{1,\dots,k-1\}})$ from equation (\ref{eq:BPeqn}) efficiently using convolutions.

For this purpose, we adopt the binary representation of a super-spin variable $X=(x^1,\dots,x^y)\in\{0,1\}^y$ (where one uses the spin-bit correspondence $\sigma=2x-1$).

We re-write the second BP equation (\ref{eq:BPeqn}) here for convenience:
\begin{align}
	\label{eq:app_BPeqn}
	\begin{aligned}
		\widehat{\eta}(X) &= \frac{1}{\widehat{z}}\sum_{X_1,\dots,X_{k-1}}\Omega(X,X_1,\dots,X_{k-1})\prod_{i=1}^{k-1}\eta_{i}(X_i)
	\end{aligned}
\end{align}
with 
\begin{align}
	\Omega(X,X_1,\dots,X_{k-1})&=\prod_{s=1}^y\mathbb{I}[x^s,x_1^s,\dots, x_y^s \ \text{not all equal}]
\end{align}
with $\mathbb{I}[A]$ the indicator function of the event $A$.
The sum $\sum_{X_1,\dots,X_{k-1}}$ contains $2^{y(k-1)}$ terms. Using convolution allows to decrease the number of operations needed in order to compute this sum.

Let $X, X'\in\{0,1\}^y$ be two super-bit configurations, and let $X\oplus X'$ be the result of the XOR operation applied component-wise.
Similarly, let $X\vee X'$ be the result of OR operation applied component-wise.

Let $h_1:\{0,1\}^y\to\mathbb{R}$, $h_2:\{0,1\}^y\to\mathbb{R}$ be a pair of discrete functions, and define the convolution as follows:
\begin{align}
	\begin{aligned}
	h_1\circledast h_2&:\{0,1\}^y\to \mathbb{R} \\
	h_1\circledast h_2(X)& = \sum_{X_1,X_2}h_1(X_1)h_2(X_2)\mathbb{I}[X=X_1\vee X_2]
	\end{aligned}
\end{align}

At fixed $X$, one can define the super-spin configurations $W_i=X\oplus X_i$, that stores the super-spin components that are satisfied by $X_i$.
One can re-write the function $\Omega$ in terms of $W_1,\dots,W_{k-1}$: 
\begin{align}
	\Omega(X,X_1,\dots,X_{k-1})&=\mathbb{I}[W_1\vee W_2 \vee \dots \vee W_{k-1}=U_y]
\end{align}
with $U_y=(1,\dots,1)$ the super-spin variable with all components equal to $1$.

With these notations, one defines recursively the functions $\{g_i:\{0,1\}^y\to \mathbb{R}\}_{i\in\{0,\dots,k-1\}}$ as follows:
\begin{align}
	\begin{aligned}
	g_i(X) &= (g_{i-1}\circledast \eta_i(X\oplus \cdot))(X)\\ &=\sum_{W_i}\sum_{X'}\eta_i(W_i\oplus  X)g_{i-1}(X')\mathbb{I}[X=X'\vee W_i]
	\end{aligned}
\end{align}
With the initialisation 
\begin{align}
	g_0(X) = \mathbb{I}[X=U_0]
\end{align}
with $U_0=(0,\dots,0)$.

One can check that:
\begin{align}
	\hat{\eta}(X) = \frac{1}{\hat{z}}g_{k-1}(U_y)
\end{align}
Computing each function $g_i$ requires a sum over two super-spin variables, therefore leading to a total cost of $2^{2y}(k-1)$ operations.

\subsubsection{Sampling super-spin configurations}
\label{app:sampling}
In this sub-section, one explains how to sample efficiently the super-spin configurations $X_1,\dots,X_{k-1}$ from the probability distribution $\nu(X_1,\dots,X_{k-1}|X,\bar{\eta}_1,\dots,\bar{\eta}_{k-1})$ given in equation (\ref{eq:distrib_nu}).
Being a distribution over $k-1$ super-spins, it is a function of $2^{y(k-1)}$ variables. However, one can avoid to compute and store the joint probability over $X_1,\dots,X_{k-1}$, by sampling them iteratively.

Using the notations introduced in appendix \ref{app:convolutions}, one write the probability distribution of $X_{k-1}$ (marginalized over the remaining super-spin configurations) $X_1,\dots,X_{k-2}$) as:
\begin{widetext}
\begin{align}
	P_\nu(X_{k-1}) = \frac{\bar{\eta}_{k-1}(X_{k-1})\sum_{Z}g_{k-2}(Z)\mathbb{I}[(Z\vee W_{k-1})=U_y]}{g_{k-1}(U_y)}
\end{align}
with the functions $g_0,\dots,g_{k-1}$ defined in the previous section, and with $W_i=X\oplus X_i$.
Similarly, the conditional probability distribution of $X_{k-i}$ given $X_{k-i+1},\dots, X_{k-1}$ is, for each $i\in\{1,\dots,k-2\}$:
\begin{align}
	P_\nu(X_{k-i}| X_{k-i+1},\dots, X_{k-1}) = \frac{\bar{\eta}_{k-i}(X_{k-i})\sum_Zg_{k-i-1}(Z)\mathbb{I}[Z\vee W_{k-i}\vee \dots \vee W_{k-1}]}{\sum_Z g_{k-i}(Z)\mathbb{I}[Z\vee W_{k-i+1}\vee \dots \vee W_{k-1}]}
\end{align}
\end{widetext}
Therefore, in order to sample the super-spin variables $X_1,\dots,X_{k-1}$, one can store the functions $g_1,\dots g_{k-1}$ computed during the update of $\bar{\eta}$, instead of the full probability distribution $\nu$.

\subsubsection{Choice of the initial condition}
\label{app:init_cond}
\begin{figure}[h!]
	\centering
	\includegraphics[width=0.5\textwidth]{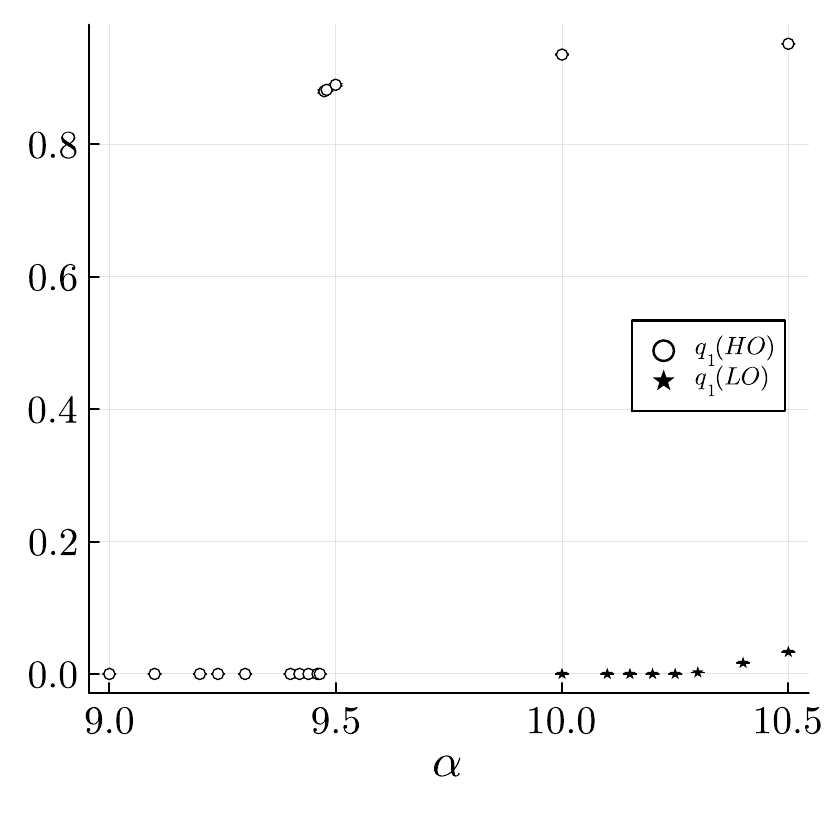}
	\caption{Intra-state overlap $q_1$ reached from two initial conditions: HO (white circles) and LO (black stars), see \ref{app:init_cond}, at $\gamma=0.01$, with population size $\mathcal{N}=10^5$. }
	\label{fig:overlap_HO_LO}
\end{figure}

In section \ref{subsubsec:phase_diagram_detailed}, we unveil the existence of (at least) two different non-trivial solutions to the 1RSB equation (\ref{eq:1RSB_x1_sym}).
In order to reach numerically these solution, we ran population dynamics with the following initialization:
\begin{align}
	P_n^{(0)}(\eta|\bar{\eta}) &= (1-\epsilon)\mathcal{P}^{\rm RS}(\bar{\eta}) + \epsilon\delta[\eta(\cdot),\delta(\cdot,U_n)]
\end{align}
with $U_n$ defined in (\ref{eq:def_U_n}). In other words, with probability $\epsilon$, the message $\eta$ is polarized on the configuration $U_n$ corresponding to the index of the probability distribution $P_n=P_{U_n}$, and with probability $1-\epsilon$ it is sampled from the RS distribution $\mathcal{P}^{\rm RS}$.

For each choice of the parameters we ran twice the population dynamics algorithm, once with $\epsilon = 1$, and once with a small value of $\epsilon > 0$ (in practice we used $\epsilon=0.01$).
We call HO, for high overlap, the initialization with $\epsilon=1$, and LO (low
overlap) the small $\epsilon$ one.

\subsubsection{Numerical determination of the thresholds}
\begin{figure*}
	\centering
	\includegraphics[width=0.9\textwidth]{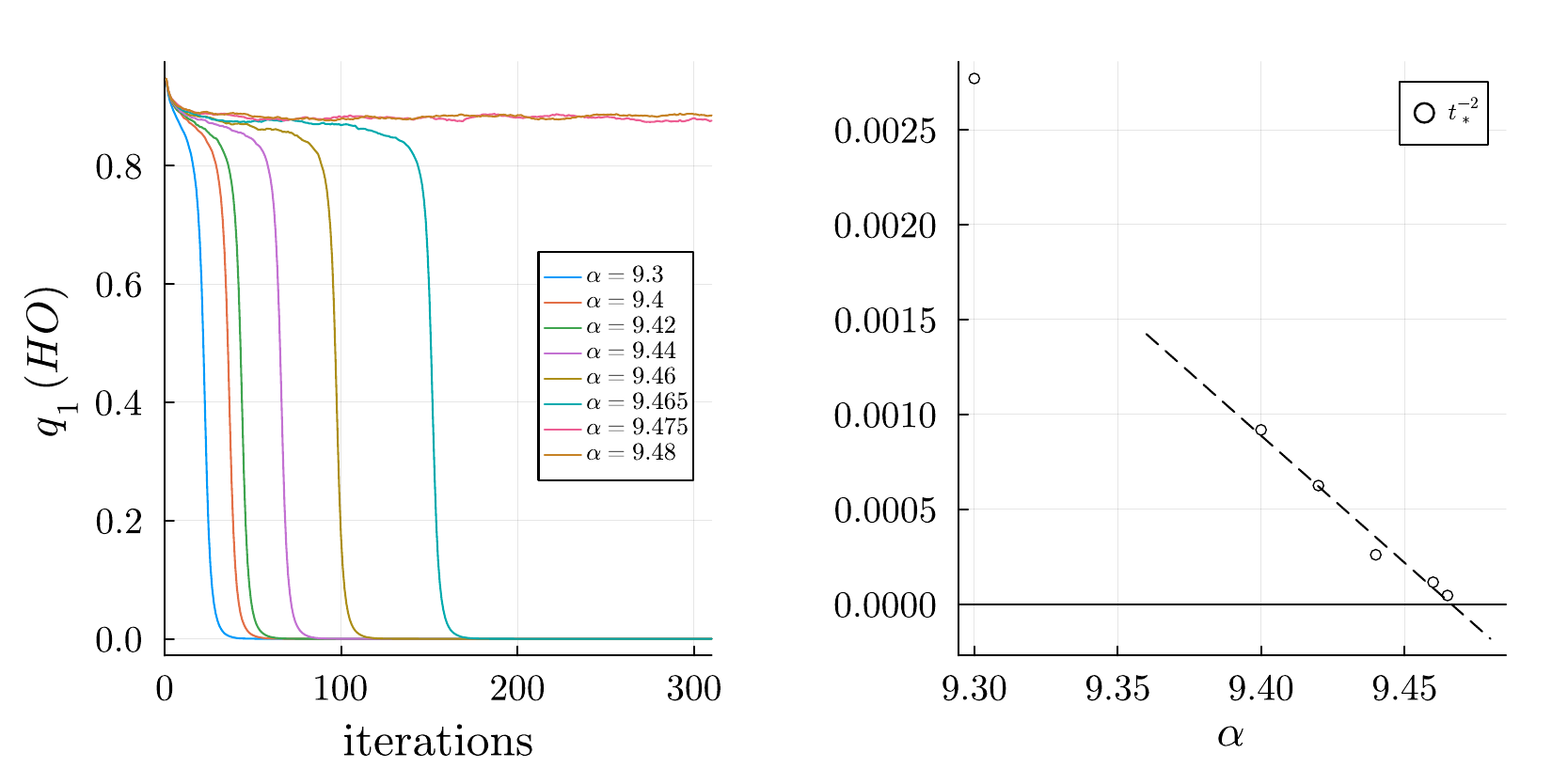}\\
	\includegraphics[width=0.9\textwidth]{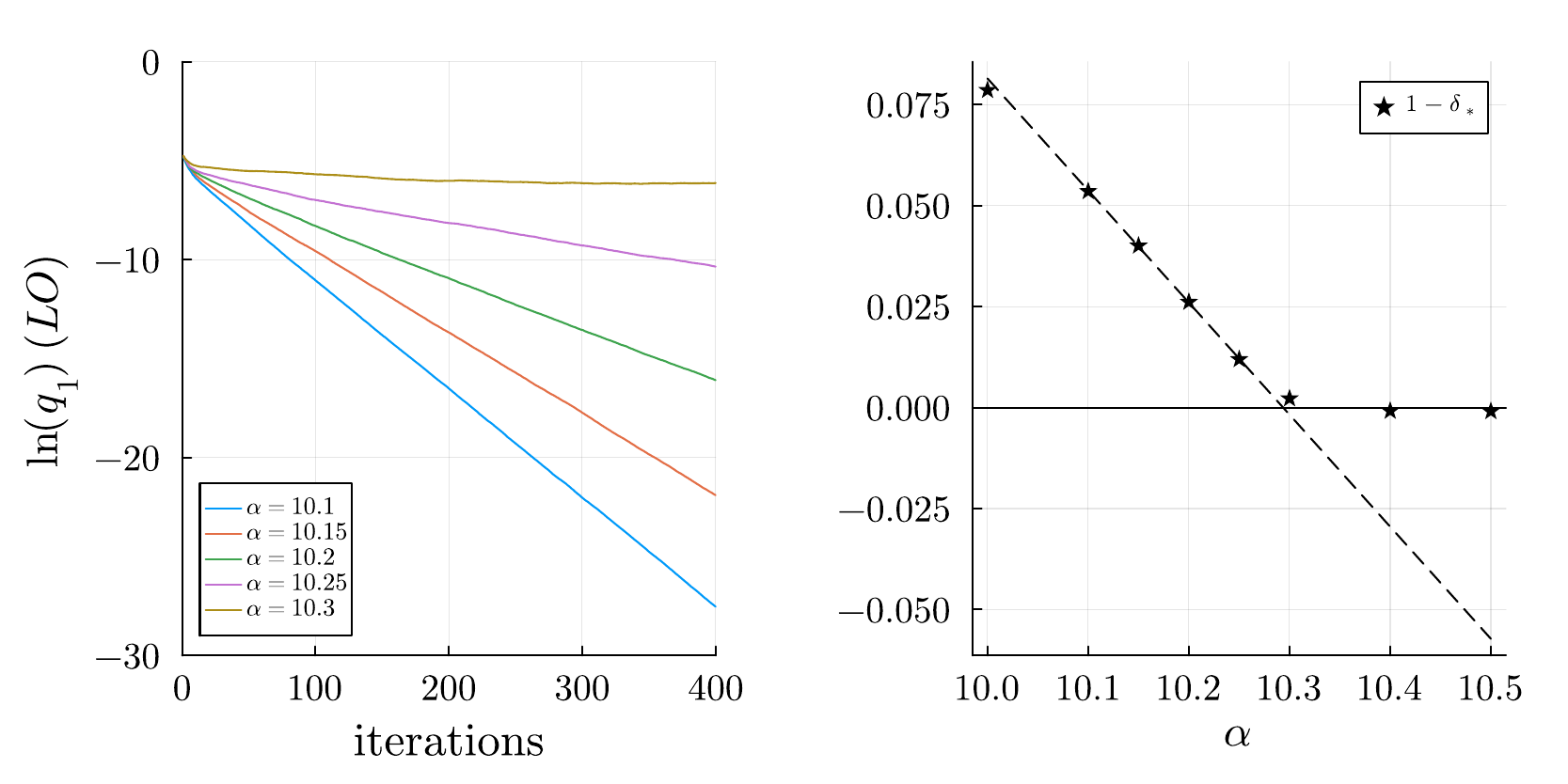}
	\caption{Numerical determination of the thresholds $\alpha_{\rm disc}(\gamma)$ (top panels), and $\alpha_{\rm KS}(\gamma)$ (bottom panels), for $\gamma=0.01$. \\
	Top: the discontinuous transition is computed from the time $t_*$ at which the overlap $q_1$ leaves the plateau and reaches $0$ (left panel). A fit using the scaling function (\ref{eq:scaling_t}) is used to compute the threshold $\alpha_{\rm disc}(\gamma=0.01)=9.461$ (right panel).\\
	Bottom: the continuous transition is computed from the intra-state overlap $q_1$ exhibiting an exponential decay for $\alpha<\alpha_{\rm KS}$ (left panel). A fit using the scaling function (\ref{eq:scaling_delta}) is used to compute the threshold $\alpha_{\rm KS}(\gamma=0.01)=10.294$.
}
	\label{fig:fit_thresholds}
\end{figure*}
In this sub-section, we give details on the numerical computation of the two thresholds $\alpha_{\rm KS}$, $\alpha_{\rm disc}$ defined in section \ref{subsec:phase_diagram}.

We recall that $\alpha_{\rm disc}$ marks the discontinuous appearance of a non-trivial solution to the 1RSB equation (\ref{eq:1RSB_x1_sym}). 
Such transition is depicted in figure \ref{fig:overlap_HO_LO}, at $\gamma=0.01$, where the intra-state overlap obtained with the HO initial condition abruptly jumps from zero to a positive value at $\alpha_{\rm disc}(\gamma=0.01)=9.461$. 
The precise numerical determination can be obtained in several ways (see e.g. \cite{BuRiSe19}, section III.) by looking at the evolution of $q_1$ under iterations (see figure \ref{fig:fit_thresholds}, top left panel). 
As $\alpha$ is growing towards $\alpha_{\rm disc}$, the time $t_*(\alpha)$ needed for $q_1^{(t)}$ to leave the plateau at positive value increases.
The time $t_*$ can be computed as the first time $q_1^{(t)}$ crosses an arbitrary value between $0$ and the plateau. Plotting $t^*$ as a function of $\alpha$ (figure \ref{fig:fit_thresholds} (top right panel)), one observes a dependence of the form:
\begin{align}
\label{eq:scaling_t}
t_* (\alpha) \simeq K (\alpha_{\rm disc} - \alpha)^{-1/2}
\quad \text{when} \quad \alpha\to\alpha_{\rm disc}
\end{align}
with $K$ a finite constant.
(as already observed in \cite{BuRiSe19}, see \cite{MoSe06} for more details on scaling functions describing the overlap evolution). The discontinuous threshold $\alpha_{\rm disc}$ is then obtained from a fit of the data using this scaling function.

The numerical determination of the continuous appearance of a non-trivial solution (or Kesten-Stigum threshold) can be obtained by studying the stability of the trivial RS solution (\ref{eq:RS_fixedpoint}) under the 1RSB equations (\ref{eq:1RSBeqn}). Here, we adopted a simpler strategy and observed that the evolution of the intra-state overlap $q_1$ under iterations is decaying exponentially to $0$ for $\alpha<\alpha_{\rm KS}$ (see figure \ref{fig:fit_thresholds}, bottom left panel), i.e.:
\begin{align}
\label{eq:scaling_delta}
q_1(t)\simeq A\delta_*^t
\end{align}
The parameter $\delta_*$ is extracted from the data $q_1(t)$ with the above scaling function, and is plotted as a function of $\alpha$ in the bottom right panel. Its value is approaching $1$ as $\alpha$ increases towards the threshold $\alpha_{\rm KS}$.
A simple linear fit of $\delta_*(\alpha)$ gives the value of the threshold $\alpha_{\rm KS}$. 
We checked that this method gives indeed the correct Kesten Stigum threshold at $\gamma=0$ (i.e. in the non-interacting case): $\alpha_{\rm KS}(k,\gamma=0) = \frac{(2^{k-1}-1)^2}{k(k-1)}=11.25$ at $k=5$.

\end{document}